\documentclass{aa}

\usepackage[utf8]{inputenc}
\usepackage[varg]{txfonts}
\usepackage{amssymb}
\usepackage{epsfig}
\usepackage{graphics}
\usepackage{amsmath}
\usepackage{color}
\usepackage{natbib}
\usepackage{hyperref}
\usepackage{gensymb}

\usepackage{aas_macros}

\usepackage{cleveref}
\usepackage{soul}

\usepackage{bm}
\usepackage{mathtools}
\usepackage{graphicx}
\usepackage{lipsum}
\usepackage{physics}
\usepackage{ulem}
\usepackage{multicol}

\definecolor{yligreen}{rgb}{0.5,0.5,0.0}

\newcommand{\rmd}{{\rm d}}
\newcommand{\rs}{{R}_{\rm S}}

\newcommand{\unit}[1]{\mbox{\boldmath $\hat{#1}$}}

\bibpunct{(}{)}{;}{a}{}{,} 

\DeclareUnicodeCharacter{00A0}{ }

\makeatletter
\def\fvec#1{\underline{\sbox\tw@{$#1$}\dp\tw@\z@\box\tw@}}
\makeatother

\begin{document}
\title{Analytical ray-tracing of synchrotron emission around accreting black holes}

\titlerunning{{\sc artpol} for synchrotron emission around accretion BHs}

\author{Alexandra~Veledina\inst{1,2}
\and Matthieu~P\'{e}lissier\inst{3} 
}

\institute{Department of Physics and Astronomy, FI-20014 University of Turku, Finland 
\email{alexandra.veledina@gmail.com}
\and Nordita, KTH Royal Institute of Technology and Stockholm University, Hannes Alfv\'{e}ns v\"{a}g 12, SE-10691 Stockholm, Sweden
\and Universit\'{e} Grenoble Alpes, CNRS, IPAG, 38000 Grenoble, France
}


\abstract{
Polarimetric images of accreting black holes encode important information about laws of strong gravity and relativistic motions of matter.
Recent advancements in instrumentation enabled such studies in two objects: supermassive black holes M87* and Sagittarius~A*.
Light coming from these sources is produced by synchrotron mechanism whose polarization is directly linked to magnetic field lines, and propagates towards the observer in a curved spacetime.
We study the distortions of the gas image by the analytical ray-tracing technique for polarized light {\sc artpol}, that is adapted for the case of synchrotron emission.
We derive analytical expressions for fast conversion of intensity/flux, polarization degree and polarization angle from the local to observer's coordinates.
We put emphasis on the non-zero matter elevation above the equatorial plane and non-circular matter motions.
Applications of the developed formalism include static polarimetric imaging of the black hole vicinity and dynamic polarimetric signatures of matter close to the compact object.
}


\maketitle

\section{Introduction}
\label{sect:intro}

Accretion onto a compact object, a black hole (BH) or a neutron star (NS), leads to a substantial energy release and excites interest to study details of energy liberation.
Thanks to the recent technological advances in radio, sub-mm and near-infrared (NIR) interferometry, we are able to probe the emission regions with micro-arc-second angular resolution. 
The highest resolution achieved in such studies, achieved by the Event Horizon Telescope (EHT), allows to produce images with size of a few Schwarzschild radii ($R_{\rm S}$). 
Two supermassive BHs --- M87* \citep{EHT2019I} and Sgr~A* \citep{EHT2022I_Sgr} --- have now been resolved down to the characteristic scales of energy release.
Emission in the EHT wavelengths is thought to be shaped by synchrotron mechanism, indicating the presence of high-energy electrons and magnetic fields of significant magnitude near the event horizon.
Additional information coming from polarimetric studies \citep{EHT2021VIII,EHT2024VII_Sgr} suggests a rather ordered topology of the field.

Both M87* and Sgr~A* are characterized by a low level of accretion, where an interplay between plasma processes and strong gravity jointly affect the observed images and average spectro-polarimetric signatures.
Disentangling between these effects is a difficult and degenerate problem.
A promising avenue to understand the role of various effects is to use the time-dependent information.

Variability studies suggest an intermittent nature of accretion onto Sgr~A*: the stable quiescent level is interrupted by incursions into bright flares. 
During the flare, the observed radio/sub-mm, NIR and X-ray fluxes increase by about an order of magnitude \citep{Baganoff2001,Genzel2003,Eckart2006_fluxes,Ponti2017,GRAVITY2020_flares,Wielgus2022}
The flares are characterized by a high linear polarization degree (PD reaching $40$\%) and rotating polarization angle (PA) over the course of the flare \citep{Eckart2006_polarimetry,Trippe2007}.
The flares occur up to a few times per day, their average duration is about an hour, the latter being consistent with the characteristic Keplerian timescale near the event horizon, although rich sub-structure morphology with timescales of minutes has been detected \citep{Dodds-Eden2009}.
Individual flares are now being resolved in both astrometric and polarimetric spaces; two flares have good simultaneous coverage in both domains \citep{GRAVITY2018,GRAVITY2020_analytical,Wielgus2022,GRAVITY2023}.

Several scenarios for the origin of flares have been discussed, either based on a hot spot orbiting the BH close to the innermost stable circular orbit \citep[ISCO,][]{Genzel2003,Wielgus2022,Vincent2023} or an expanding blob of plasma ejected from the BH vicinity \citep{Yusef-Zadeh2006,Vincent2014,Aimar2023}.
More sophisticated models have also been considered, such as an orbiting complex emission pattern, e.g. formed by the local perturbations or by the interplay between the outflow and surrounding disk \citep{Matsumoto2020}.
Finally, it remains elusive whether all flares can be attributed to a single mechanism, or, rather, similar flares can be caused by different physical phenomena.
In this context, it is important to note the recent change of statistical properties of flaring activity in Sgr A*, likely owing to the close approaches by S0-2 stars and G2 cloud \citep{Do2009,Do2019}.
 
Intrinsic spectra and polarimetric properties of the hot plasma near the event horizon are altered on the way to the distant observed by the relativistic aberration, light bending and frame dragging effects (\citealt{ConnorsStark1977}; \mbox{\citealt{StarkConnors1977}}; \citealt{Pineault1977pol_schw,PineaultRoeder1977kerr_analyt,ConnorsPiranStark1980}).
While the PD is an invariant quantity, the polarization signatures of each point/element in the accretion disk is expected to be different because of the PA rotation and alteration of the viewing angle; both of these effects depend on the location of the element in the disk, BH spin and observer inclination.
To understand the intrinsic properties of flares, as well as the geometry of the quiescent-level accretion onto Sgr~A* and M87*, it is important to distinguish between the effects of strong gravity, radiative transfer and plasma processes.
Calculations of these effects often involve parallel transport of the Stokes vector along the null geodesics, e.g. via the conservation of the Walker-Penrose \citeyear{WalkerPenrose1970} constant.
The geodesics can in turn be computed using the ray-tracing technique \citep[often used, e.g.,][]{Dexter2009,Vos2022}.
However, this straightforward routine is often non-intuitive with respect to the prediction of parameters from the observed properties; it is also rather computationally expensive.
 
Recently, a number of semi-analytical methods have been developed to serve the purpose of computationally fast studies of the effects of model parameters on the observed images, polarimetry and variability.
Polarimetric images of the M87* surroundings have been obtained \citep{Narayan2021} by evaluating the Walker-Penrose constant for an approximate light bending formula of \citet{Beloborodov2002}.
An extension to this model for the case of a Kerr BH was also considered \citep{Gelles2021}.
Furthermore, the effects of relativistic aberration have been isolated by considering the polarimetric signatures of orbiting hot spots in Minkowski space \citep{Vincent2023}.
The aforementioned methods consider motion of matter in the equatorial plane; more recently, images of the vertically-extended source have also been computed \citep{Chang2024}.
The effects of the non-zero matter elevation on polarimetric properties have not been given detailed consideration.

In this work we study the importance of effects of the non-equatorial and non-circular motion on the observed imaging and polarimetric characteristics.
We utilize the approach of analytical ray tracing for spectro-polarimetric ({\sc artpol}) calculations and give explicit expressions for the PA rotation.
The method is based on calculation of rotation of PA along the  photon trajectory using the higher-order approximation to the light bending formula \citep{Poutanen2020bending}.
It was proved to be robust in the context of the X-ray polarization formed in the atmospheres of accreting NSs \citep{Loktev20}, accretion disks around Schwarzschild \citep{Loktev2022} and Kerr \citep{Loktev2024} BHs.
We extend the previous calculations for (i) the cases of polarized synchrotron emission, (ii) non-circular motion of matter and (iii) non-zero matter elevation above the equatorial plane and non-zero thickness of the disk.
{  We consider the case of Schwarzschild metric, however, general expressions derived in this work can be used also in the case of Kerr metric.}

The first extension introduces an extra freedom in terms of magnetic field orientation and adds azimuthal dependence of local quantities.
The second extension alters the azimuthal angles where Doppler effects are important.
Addition of the third effect plays an important role in comparison to the observations.

The paper is organized as follows.
In Sect.~\ref{sect:setup} we describe the geometry, introduce the coordinate system and auxiliary quantities.
We also give explicit expression for the rotation of PA, as a function of azimuth, caused by general, special relativity, as well as by the changes of local magnetic field orientation.
Vector form for calculation of the aforementioned rotation angles, detailed comparison to the previous methods and approximations are given in Appendix.
In Sect.~\ref{sect:results} we show the examples of calculations with the emphasis on the difference between the resulting quantities with the new features included.
In Sect.~\ref{sect:applications} we consider applications of the developed approach to the BH images and spectro-polarimetric properties of flares.
We also detail the difference between our results and previous semi-analytical models.
In Sect.~\ref{sect:summary} we summarize our findings.

\begin{figure*} 
\centering
\includegraphics[width=0.75\linewidth]{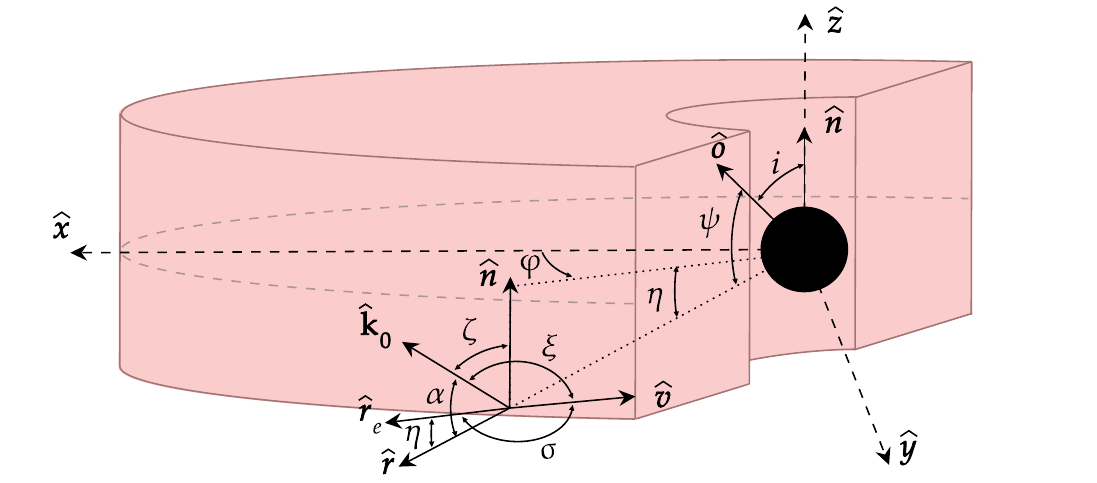}
\caption{Geometry of the accreting matter around the BH.
The fluid element is described by the radius vector $\unit{r}$ and velocity $\unit{v}$. Its trajectory is shifted from the mid-plane of the accretion disk by the angle $\eta$. The observer is located along the direction of $\unit{o}$, making an angle $i$ with the disk normal $\unit{n}$.
}
\label{fig:geom}
\end{figure*}

\section{Model setup}
\label{sect:setup}

In this section we introduce the formalism describing the accretion geometry and transformation of the basic vectors affected by the strong gravity and fast motions of matter.
We note that the formalism partly repeats the description given in \citet{Loktev2022,Loktev2024}, however, we give the full description here for completeness.
Motivated by the earlier studies of M87* and Sgr A* images and flares showing secondary effect of BH spin on the observed characteristics \citep[e.g.,][]{Gelles2021}, we consider the case of a non-spinning BH.

\subsection{Accretion geometry and observed flux}
\label{sect:accrdisc}

We consider an accreting matter around the BH in Schwarzschild metric.
The matter can either constitute a compact blob (hot spot), or form an extended structure (accretion disk or its narrow ring).
In both cases we identify the basic plane related to the equatorial (mid-)plane of the accretion disk and introduce the Cartesian coordinates related to this plane (see Fig.~\ref{fig:geom}).
We choose $z$-axis to coincide with the normal to the disk plane $\unit{n}$.
The $x$-axis lies along the projection of the line of sight on the disk.
The unit vector $\unit{o}$ points towards the observer and
the radius vector of the considered point in the disk/hot spot is $\bm{r}$ (with the corresponding unit vector $\unit{r}$) and the unit azimuthal vector is $\unit{\varphi}$. 
These vectors have the following coordinates:
\begin{eqnarray}\label{eq:vectors_cart}
    \unit{n} & = & (0, 0, 1), \\
    \unit{o} & = & (\sin i, 0, \cos i), \\
    \unit{r}   & = & (\cos\varphi\cos\eta, \sin\varphi\cos\eta, \sin\eta), \\
    \unit{\varphi} & = & (-\sin\varphi, \cos\varphi, 0),
\end{eqnarray}
where $\eta$ is the angle between the radius-vector $\unit{r}$ and the equatorial plane, $\varphi$ is the azimuthal angle of the projection of $\unit{r}$ to the $x$-$y$ plane, measured from the $x$-axis and $i$ is the inclination of the disk plane to the line of sight.
The vectors $\unit{r}$ and $\unit{o}$ form an angle $\psi$:
\begin{equation}
    \cos\psi \equiv \unit{r} \cdot \unit{o} = \sin i \cos\varphi \cos\eta + \cos i \sin\eta.
\end{equation}
Due to the effects of the light bending, the photons reaching the observer along $\unit{o}$ are leaving the disk along the other direction, which we denote as $\unit{k}_0$.
Because the photon trajectories lie in a plane in Schwarzschild metric, the wave vector $\unit{k}_0$ can be described by a linear combination of the radius-vector and the observer direction:
\begin{equation}
  \unit{k}_0 = [\sin\alpha\ \unit{o} + \sin(\psi-\alpha) \unit{r}]/\sin\psi,
\end{equation}
where $\alpha$ is the angle between the radius vector and direction towards the observer,
\begin{equation}
\cos\alpha  \equiv \unit{r} \cdot \unit{k}_0.
\end{equation}  
The angle $\alpha$ is related to $\psi$ through the light-bending integral or an approximate light bending formula \citep{Pechenick1983,Beloborodov2002,Poutanen2020bending}, introduced explicitly below.
The wave vector makes an angle $\zeta$ with the disk normal
\begin{equation}\label{eq:coszeta}
    \cos\zeta \equiv \unit{n} \cdot \unit{k}_0 = \frac{\sin\alpha \cos i + \sin(\psi - \alpha) \sin\eta}{\sin\psi}.
\end{equation}
 
Plasma orbits around the BH with Keplerian velocity and additionally gradually moves towards the compact object.
The unit velocity vector is thus not purely parallel to the azimuthal vector. 
This deviation is described by the angle $\sigma$ with respect to the equatorial radius vector $\unit{r}_{\rm e}=(\cos\varphi, \sin\varphi, 0)$:
\begin{equation}
    \cos\sigma  \equiv \unit{r}_{\rm e} \cdot \unit{v}.
\end{equation}  
We note that we do not consider vertical motion of matter (along the $z$-axis).
The unit velocity vector is described by
\begin{equation} 
    \unit{v} = (\cos(\varphi + \sigma), \sin(\varphi + \sigma), 0).
\end{equation}
Purely Keplerian rotation can be recovered by choosing $\sigma=90\degr$ or $270\degr$; matter moves towards the BH for $90\degr < \sigma < 270\degr$.
For the magnitude of dimensionless velocity, $\beta=v/c$, relative to the distant observer, we adopt the relation of \citet{Luminet1979}.
In this work, we consider the case where the matter above the disk at height $h=r\sin\eta$, where $r=R/R_{\rm S}$ is the circumferential radius measured in units of Schwarzschild radii $R_{\rm S}=2GM/c^2$ ($M$ is the mass of the compact object) and $\eta\neq0$, moves with Keplerian velocity corresponding to its equatorial radius, $r_{\rm e}=\sqrt{r^2-h^2}$.
We have
\begin{equation}\label{eq:beta}
    \beta = \frac{1}{\sqrt{2(r_{\rm e}-1)}}
\end{equation}
and the Lorentz factor is then 
\begin{equation} 
 \gamma \equiv \frac{1}{\sqrt{1-\beta^2}} = \sqrt{ \frac{r_{\rm e}-1}{r_{\rm e}-1.5} }.
\end{equation}
The angle between the photon momentum and velocity vector is expressed as
\begin{equation}\label{eq:cosxi}
    \cos\xi \equiv \unit{v} \cdot \unit{k}_0 = 
    \frac{\sin\alpha}{\sin\psi} \sin i \cos(\varphi+\sigma) + \frac{\sin(\psi-\alpha)}{\sin\psi} \cos\eta \cos\sigma.
\end{equation}

Fast motions of matter close to the BH cause the photon trajectory to be altered by relativistic aberration.
To account for this effect in the observed polarization properties, we consider quantities in a local fluid co-moving frame.
We denote quantities measured in this frame with primes.
The photon vector in the co-moving frame is
\begin{equation}
    \unit{k}^{\prime}_0 = 
    \delta [\unit{k}_0 - \gamma \beta \unit{v} + (\gamma-1) \unit{v} (\unit{v} \cdot \unit{k}_0)],
    \label{eq:kprime}
\end{equation}
where $\delta$ is the Doppler factor
\begin{equation}\label{eq:factors}
 \delta \equiv \frac{1}{\gamma(1-\beta \unit{v}\cdot\unit{k}_0)} = \frac{1}{\gamma(1-\beta \cos\xi)}.
\end{equation}
In the fluid frame, the photon makes angle $\xi'$ with the velocity vector
\begin{equation}
    \cos\xi' \equiv \unit{v} \cdot \unit{k}^{\prime}_0 = \frac{\cos\xi - \beta}{1 - \beta\cos\xi} = \gamma \delta (\cos\xi - \beta).
\end{equation}
The angle between the photon momentum and the local normal in the disk is
\begin{equation}\label{eq:coszetap}
    \cos\zeta' \equiv \unit{n} \cdot \unit{k}^{\prime}_0 = \delta \cos\zeta.
\end{equation}

To account for the light bending effects and to transform the local emission and polarization to the observer frame, we need a relation between the angle $\alpha$ of the outgoing photon and the observer direction $\psi$.
We use the approximate relation \citep{Poutanen2020bending}
\begin{equation}
    \cos\alpha \approx 1- (1-1/r)y\left\{ 1+\frac{y^2}{112r^2}- \frac{{\rm ey}}{100r} \left[ \ln\left(1-\frac{y}{2}\right) +\frac{y}{2} \right] \right\},
\end{equation}
where $y = 1-\cos\psi$. 
This relation gives accurate results given the photon trajectories lie in the plane, that is strictly satisfied in Schwarzschild metric.
Transition from Schwarzschild to Kerr metric can be made by accounting for two additional effects: (i) change of the orbital velocity profile and (ii) additional twist of light trajectories caused by the frame dragging effects. 
While the former step is straight-forward as proper velocity profile is known, the latter effect is not accounted for by the present light bending approximation.
However, the second effect introduces deviations only for the trajectories that lie very close to the spinning BH, $<3\rs$. Significant errors can be expected for the trajectories originating from the parts of the disk behind the BH ($\varphi\sim180\degr$) and a combination of high spin parameter $a\gtrsim0.8$ and high inclination ($i\gtrsim60\degr$).
At the same time, the integrated PD and PA in these cases show modest deviations ($0.5$\% and $7\degr$), becoming comparable to the observational uncertainties \citep{Loktev2024}.
For the cases of prograde spins $a<0.8$, retrograde spins up to $a=-1$ and/or orbits at $\sim3\rs$ or further (comparable to the scales of the BH shadow probed by EHT), the current light bending approximation leads to a precision higher than the observational errors.

The accretion flow is assumed to be optically thin to its own synchrotron emission and its spectral energy distribution follows the power-law dependence $I_{E}\propto E^{-\alpha_{E}}$.
This means that the observer sees the whole volume of the flow, rather than its surface emission.
To treat this, we split the accretion flow in surfaces in cylindrical coordinates, with the $z$-axis being perpendicular to the equatorial plane of the accretion disk, and compute the total flux from each surface parallel to the equatorial plane by further splitting it into segments in radial and azimuthal direction.
The observed flux of each such segment depends on the solid angle it occupies on the sky \citep[see more details in][]{PB06,Loktev2022} 
\begin{equation}
    \rmd \Omega =\frac{ \rmd S \cos\zeta}{D^2}   {\cal L} = \frac{\rs^2}{D^2} \frac{ r \, \rmd r \, \rmd \varphi }{\sqrt{1-1/r}} {\cal L} \cos\zeta,
\end{equation}
where factor $\rmd S \cos\zeta$ comes from the projection of the segment, $D$ is the distance and ${\cal L}$ is the lensing factor \citep{Beloborodov2002,Poutanen2020bending}
\begin{eqnarray}\label{eq:lensing}
    {\cal L} &\equiv& \frac{1}{1-1/r} \frac{\rmd \cos\alpha}{\rmd \cos\psi} \approx \\
    & \approx & 1 + \frac{3y^2}{112r^2} - \frac{{\rm e}y}{100 r} \left[ 2 \ln{\left(1-\frac{y}{2}\right)+y\frac{1-3y/4}{1-y/2}} \right].
\end{eqnarray}
The segment area projected onto the plane orthogonal to the photon propagation is Lorentz invariant: $\rmd S' \cos\zeta' = \rmd S \cos\zeta$. 
Hence, the observed solid angle of can be expressed through the area of the disk segment as
\begin{equation}
    \rmd \Omega = \frac{ \delta \rmd S' \cos\zeta}{D^2} {\cal L}.
\end{equation}
The latter expression will be used for computing the fluxes from the moving spots (blobs), for which the intrinsic area $\rmd S'$ remains constant during the orbital motion.
This area can be expressed through the comoving coordinates $(r',\varphi')$ as \citep[see also][]{Nattila2018,Bogdanov2019,Suleimanov2020}
\begin{equation}
    \rmd S' = \frac{\gamma \rs^2 r'_{\rm e} \rmd r'_{\rm e} \rmd \varphi'}{\sqrt{1-1/r}}.
\end{equation}

\subsection{Transformation of polarized synchrotron emission}

The magnetic field is described in the co-moving frame, where it has radial, azimuthal and vertical components. 
It is convenient to define the field through the components lying in the equatorial plane $B_{\rm eq}$ and along the disk normal $B_z$.
The equatorial field is further divided into the component lying along the gas motion and orthogonal to that, described by the angle $\phi'$ of the magnetic field direction relative to the gas motion.
In the Cartesian coordinates of the co-moving frame, the magnetic field vector is given by
\begin{equation}
    {\bm B}' = B_{\rm eq} \left[ \cos{\phi'} \unit{v} + \sin{\phi'} (\unit{n} \cross \unit{v}) \right] + B_z \unit{n}.
\end{equation} 
The unit vector $\unit{B}'\equiv {\bm B}'/|{\bm B}'|$ has components $b_{\rm eq}$ and $b_z$.
Intensity and polarization of photon depends on the angle between the magnetic field and photon direction
\begin{equation}
    \cos\theta' \equiv \unit{B}' \cdot \unit{k}'_0 = 
    b_{\rm eq} (\cos{\phi'} \cos\xi' - 
    \delta \Tilde{b} \sin{\phi'} ) +  \delta b_z \cos{\zeta}, 
\end{equation}
where
\begin{equation}\label{eq:tildeb}
    \Tilde{b} = \frac{\sin\alpha}{\sin\psi} \sin i \sin(\varphi+\sigma) + \frac{\sin(\psi-\alpha)}{\sin\psi} \cos\eta \sin\sigma.
\end{equation}

The radiation field in the co-moving frame can be generally described by the Stokes vector 
\begin{equation}
 \bm{I}'_{E'}(\theta') \equiv \left[ \begin{array}{c} 
I_{E'}' \\ Q_{E'}' \\ U_{E'}' 
\end{array} \right] = I_{E'}'
\left[ \begin{array}{c} 
1 \\ P_{\rm s}(E') \cos 2\chi_0 \\ P_{\rm s}(E') \sin 2\chi_0
\end{array} \right] , 
\end{equation}
where $I_{E'}'$ is the local rest-frame specific intensity, $P(E')$ corresponds to the PD in this frame and $\chi_0$ gives the PA relative to the chosen axis.
A natural choice for synchrotron emission is the axis going along the magnetic field vector.
The basis vectors are
\begin{equation}
\label{eq:polbas_on}
\unit{e}_1 = \frac{\unit{B'}-\cos{\theta'}\  \unit{k}^{\prime}_0}{\sin{\theta'}},\qquad
\unit{e}_2 = \frac{\unit{k}^{\prime}_0 \times \unit{B'}}{\sin{\theta'}} .
\end{equation}
In this basis, $\chi_0=90\degr$ and synchrotron emission and polarization can be expressed as
\begin{equation}
 \bm{I}'_{E'}(\theta')  = c_0 E'^{-\alpha_E} l_{\rm ph}' \left( B \sin\theta' \right)^{1+\alpha_{E}}
\left[ \begin{array}{c} 
1 \\ - P_{\rm s} \\ 0 
\end{array} \right] , 
\end{equation}
where $c_0$ is a constant (e.g. eq.~6.36 in \citealt{RL79}), $l_{\rm ph}'$ is the photon path length through the synchrotron-emitting medium (as measured in the comoving frame), $P_{\rm s}=(p+1)/(p+7/3)$ gives the polarization degree (PD) of the power-law electrons with index $p=2\alpha_{E}+1$ \citep{Ginzburg1965}.
The typical value considered for the optically-thin synchrotron emission is $\alpha_{E}=0.7$, while the observations of bright flares in Sgr~A$^*$ tend to give somewhat steeper slope, $\alpha_{E}\approx1$ \citep{Gillessen2006}. 
The latter value was used in the modeling of the Sgr~A$^*$ and M87$^*$ \citep{Narayan2021,Vincent2023}, and we adopt this value for calculations presented in this work.
The advantage of such representation of the observed Stokes parameters is that it explicitly distinguishes the polarization angle (PA) from PD.
PD is Lorentz invariant, hence the transformation of polarization signatures reduces to the tracing of the changes of the photon electric field orientation (as a result of aberration and light bending) along the photon trajectory.

The intensity of the steady matter flow in the comoving frame is related to the observed intensity as
\begin{equation}
    I_{E} = g^{3} I_{E'}' \propto g^{3+\alpha_E} E^{-\alpha_E},
\end{equation}
where $g \equiv E/E' =\delta \sqrt{1-1/r}$ is the total redshift factor \citep{Luminet1979,CS89} and the latter expression corresponds to the power-law spectrum $I_{E'}' \propto E'^{-\alpha_E}$.
The photon path can be expressed as
\begin{equation}
    l_{\rm ph}'=g l_{\rm ph} = \frac{h_0}{\hat{k}'^{(z)}_0} = \frac{h_0}{\cos\zeta'},
\end{equation}
where $h_0$ is the characteristic vertical thickness of each synchrotron-emitting layer in the disk, as seen in the observer's frame, $\hat{k}'^{(z)}_0$ is the $z$ component of the photon momentum vector in the comoving frame.

The observed flux from a segment of the disk is related to the observed intensity of radiation as
\begin{equation}
  \rmd F_E = I_{E} \rmd \Omega = \delta I_{E} {\cal L} \cos\zeta \frac{\rmd S'}{D^2}.
\end{equation}
Stokes parameters (in flux units) of each layer in the disk is obtained by integrating over the equatorial radii and azimuthal coordinates. 
\begin{eqnarray}\label{eq:flux_stokes}
    \bm{F}_E &\equiv& \left[ \begin{array}{c} 
F_{I} (E) \\ F_Q(E) \\ F_U(E) 
\end{array} \right] = \int \rmd \Omega \left[ \begin{array}{c} 
I_{E} \\ Q_{E} \\ U_{E} 
\end{array} \right] = \\
  & = & \int \rmd \Omega \, g^3 {\bf M}(r,\varphi) \bm{I}'_{E'}(\theta') = \nonumber \\
  & = &   \int \frac{\rmd S'}{D^2} \, g^3 \delta {\cal L} \cos\zeta\  {\bf M}(r,\varphi) \bm{I}'_{E'}(\theta'), \nonumber
\end{eqnarray}
where the transformation from the vector $\bm{I}'_{E'}(\theta')$ to the observed Stokes vector is done through the rotation Mueller matrix
\begin{equation}\label{eq:pol-rot-matr}
    {\bf M}(r,\varphi) = \begin{bmatrix}
        1 & 0 & 0 \\
        0 & \cos2\chi^{\rm tot} & -\sin2\chi^{\rm tot} \\
        0 & \sin2\chi^{\rm tot} & \cos2\chi^{\rm tot}
    \end{bmatrix}.
\end{equation}
Here, $\chi^{\rm tot}$ corresponds to the total rotation of the polarization (Stokes) vector from the comoving frame to the observer's frame.
For the synchrotron emission of the power-law electrons, the  expression in Eq.~\eqref{eq:flux_stokes} can be simplified to
\begin{eqnarray}\label{eq:flux}
     \bm{F}_E =  f_{E,0} \int \rmd \Omega \, \
     \frac{g^{3+\alpha_E}}{\delta \cos\zeta} \left( \sin\theta' \right)^{1+\alpha_{E}} 
\left[ \begin{array}{c} 
1 \\ - P_{\rm s} \cos2\chi^{\rm tot} \\ - P_{\rm s} \sin2\chi^{\rm tot}
\end{array} \right] = \nonumber \\ 
  =  f_{E,0} \iint \frac{\gamma \rs^2 r'_{\rm e}\rmd r'_{\rm e}\rmd\varphi'}{\sqrt{1-1/r}}  \, 
   g^{3+\alpha_E} {\cal L} \left( \sin\theta' \right)^{1+\alpha_{E}}
\left[ \begin{array}{c} 
1 \\ - P_{\rm s} \cos2\chi^{\rm tot} \\ - P_{\rm s} \sin2\chi^{\rm tot}
\end{array} \right] ,
\end{eqnarray}
where $\displaystyle f_{E,0}= c_0 h_0 E^{-\alpha_E}B^{1+\alpha_{E}}$.
In the latter representation we have preserved the expression of $\rmd S'$, as it is most useful for moving spots, while for former expression is useful for calculations of the synchrotron-emitting disk image.
Below we also use the relative Stokes parameters, obtained as $q=F_Q(E)/F_I(E)$ and $u=F_U(E)/F_I(E)$.
Loktev2024
The developed formalism can be used to construct the spectro-polarimetric images \citep[see also][]{Loktev2022,Loktev2024}.
For this, we introduce the Cartesian coordinates $X$ and $Y$ in the plane of the sky, scaled to units of $R_{\rm S}$, as
\begin{equation}\label{eq:artimage}
    \begin{bmatrix}
        X \\ Y   
    \end{bmatrix}
 =  b  
    \begin{bmatrix}
      -\sin\Phi \\
       \cos\Phi
    \end{bmatrix} 
= \frac{r}{\sqrt{1-1/r}} \frac{\sin\alpha}{\sin\psi}
    \begin{bmatrix}
       \sin\varphi \cos\eta \\
       \sin i\sin\eta -\cos i \cos\varphi \cos\eta
     \end{bmatrix}.
\end{equation}
Here $\Phi$ corresponds to the position angle of the point in the plane of the sky, measured counterclockwise from the projection of the disk axis on the sky and $b=r\sin\alpha/\sqrt{1-1/r}$ corresponds to the impact parameter.

\begin{figure*}
\centering
\includegraphics[width=\textwidth]{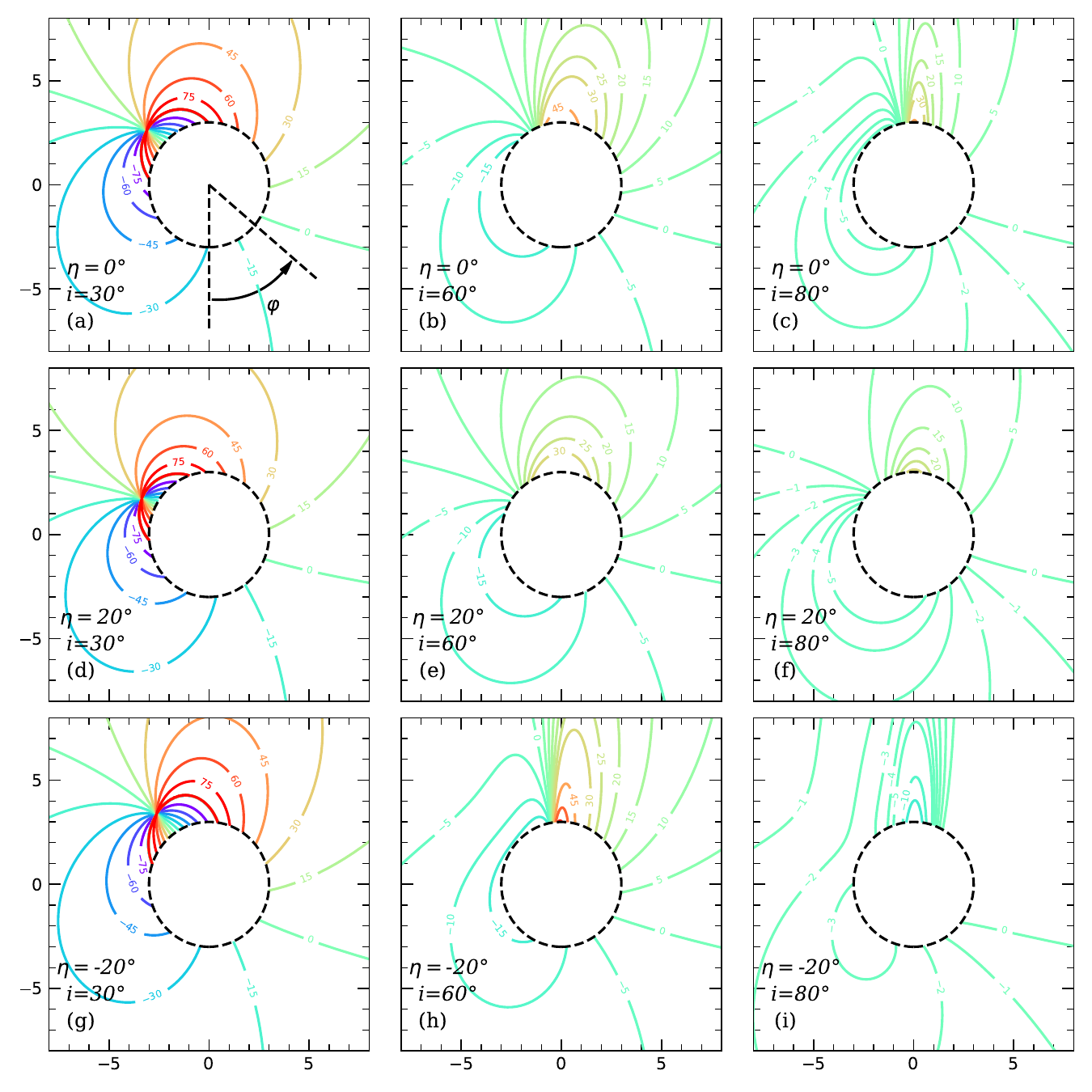}
\caption{Contours of PA rotation ($\chi^{\rm SR}+\chi^{\rm GR}$) for different $i=30\degr$, $60\degr$ and $80\degr$ and $\eta=0, \pm20\degr$ in the disk coordinates $(r,\varphi)$, where $r$ is measured in units of Schwarzschild radii.
Azimuth increases in the counterclockwise direction, as shown by the arrow in panel (a).
Numbers on the contours correspond to the rotation angle (in degrees). 
Thick red line without a mark corresponds to $\pm90\degr$ rotation.
Matter is assumed to have purely circular Keplerian velocity in the counterclockwise direction ($\sigma=90\degr$).
Dashed line indicates the location of the ISCO. 
Contours corresponding to $\eta=0$ (upper row) match those shown in \citet{Loktev2022}.
The effects of matter elevation above(/below) the orbital plane are clearly visible when comparing to the case of in-plane rotation. 
If matter is located in front of the BH (middle row), for the same radius from the BH the GR and SR effects appear less pronounced.
}
\label{fig:chiGRSR_contours}
\end{figure*}

\begin{table*}[h!]
\caption{Parameters of the models: inclination ($i$), velocity direction ($\sigma$), magnetic field ($b_{\rm eq}$, $\phi'$, $b_z$), elevation ($\eta$) and equatorial radius ($r_{\rm e}$).}          
\label{table:polar}     
\centering  
\begin{tabular}{cccccccc}    
\hline\hline               
Figure & $i$ & $\sigma$ & $b_{\rm eq}$ & $\phi'$ & $b_z$ & $\eta$ & $r_{\rm e}$\\    
  &  (deg) &  (deg) &   &  (deg)  & & (deg) \\  
\hline                       
   Fig.~\ref{fig:PAinitial}a  & $30$ & $90$ & $1$ & $90$ & $0$ & $0$ & $3,5,15,50$\\ 
   Fig.~\ref{fig:PAinitial}b  & $60$ & $90$ & $1$ & $90$ & $0$ & $0$ & $3,5,15,50$ \\ 
   Fig.~\ref{fig:PAinitial}c  & $80$ & $90$ & $1$ & $90$ & $0$ & $0$ & $3,5,15,50$ \\ 
   Fig.~\ref{fig:PAinitial}d  & $30$ & $90$ & $1$ & $0$ & $0$ & $0$ & $3,5,15,50$ \\ 
   Fig.~\ref{fig:PAinitial}e  & $60$ & $90$ & $1$ & $0$ & $0$ & $0$ & $3,5,15,50$ \\ 
   Fig.~\ref{fig:PAinitial}f  & $80$ & $90$ & $1$ & $0$ & $0$ & $0$ & $3,5,15,50$\\ 
 Fig.~\ref{fig:chitot_eta20}a & $30$ & $90$ & $1$ & $90$ & $0$ & $20$ & $3,5,15,50$ \\ 
 Fig.~\ref{fig:chitot_eta20}b & $60$ & $90$ & $1$ & $90$ & $0$ & $20$ & $3,5,15,50$ \\ 
 Fig.~\ref{fig:chitot_eta20}c & $80$ & $90$ & $1$ & $90$ & $0$ & $20$ & $3,5,15,50$ \\ 
 Fig.~\ref{fig:chitot_eta20}d & $30$ & $90$ & $1$ & $0$ & $0$ & $20$ & $3,5,15,50$ \\ 
 Fig.~\ref{fig:chitot_eta20}e & $60$ & $90$ & $1$ & $0$ & $0$ & $20$ & $3,5,15,50$ \\ 
 Fig.~\ref{fig:chitot_eta20}f & $80$ & $90$ & $1$ & $0$ & $0$ & $20$ & $3,5,15,50$ \\ 
 Fig.~\ref{fig:chitot_eta20}g & $30$ & $90$ & $0$ & $0$ & $1$ & $20$ & $3,5,15,50$ \\ 
 Fig.~\ref{fig:chitot_eta20}h & $60$ & $90$ & $0$ & $0$ & $1$ & $20$ & $3,5,15,50$ \\ 
 Fig.~\ref{fig:chitot_eta20}i & $80$ & $90$ & $0$ & $0$ & $1$ & $20$ & $3,5,15,50$ \\ 
 Fig.~\ref{fig:diff_chitot_eta20}a & $30$ & $90$ & $1$ & $90$ & $0$ & $20$ & $3,5,15,50$ \\ 
 Fig.~\ref{fig:diff_chitot_eta20}b & $60$ & $90$ & $1$ & $90$ & $0$ & $20$ & $3,5,15,50$ \\ 
 Fig.~\ref{fig:diff_chitot_eta20}c & $80$ & $90$ & $1$ & $90$ & $0$ & $20$ & $3,5,15,50$ \\ 
 Fig.~\ref{fig:diff_chitot_eta20}d & $30$ & $90$ & $1$ & $0$ & $0$ & $20$ & $3,5,15,50$ \\ 
 Fig.~\ref{fig:diff_chitot_eta20}e & $60$ & $90$ & $1$ & $0$ & $0$ & $20$ & $3,5,15,50$ \\ 
 Fig.~\ref{fig:diff_chitot_eta20}f & $80$ & $90$ & $1$ & $0$ & $0$ & $20$ & $3,5,15,50$ \\ 
 Fig.~\ref{fig:diff_chitot_eta20}g & $30$ & $90$ & $0$ & $0$ & $1$ & $20$ & $3,5,15,50$ \\ 
 Fig.~\ref{fig:diff_chitot_eta20}h & $60$ & $90$ & $0$ & $0$ & $1$ & $20$ & $3,5,15,50$ \\ 
 Fig.~\ref{fig:diff_chitot_eta20}i & $80$ & $90$ & $0$ & $0$ & $1$ & $20$ & $3,5,15,50$ \\ 
   Fig.~\ref{fig:chi_sigma}a  & $30$ & $90,120,150,180$ & $1$ & $90$ & $0$ & $0$ & $5$ \\ 
   Fig.~\ref{fig:chi_sigma}b  & $60$ & $90,120,150,180$ & $1$ & $90$ & $0$ & $0$ & $5$ \\ 
   Fig.~\ref{fig:chi_sigma}c  & $80$ & $90,120,150,180$ & $1$ & $90$ & $0$ & $0$ & $5$ \\ 
   Fig.~\ref{fig:chi_sigma}d  & $30$ & $90,120,150,180$ & $1$ & $0$ & $0$ & $0$ & $5$ \\ 
   Fig.~\ref{fig:chi_sigma}e  & $60$ & $90,120,150,180$ & $1$ & $0$ & $0$ & $0$ & $5$ \\ 
   Fig.~\ref{fig:chi_sigma}f  & $80$ & $90,120,150,180$ & $1$ & $0$ & $0$ & $0$ & $5$ \\ 
   Fig.~\ref{fig:chi_sigma}g  & $30$ & $90,120,150,180$ & $0$ & $0$ & $1$ & $0$ & $5$ \\ 
   Fig.~\ref{fig:chi_sigma}h  & $60$ & $90,120,150,180$ & $0$ & $0$ & $1$ & $0$ & $5$ \\ 
   Fig.~\ref{fig:chi_sigma}i  & $80$ & $90,120,150,180$ & $0$ & $0$ & $1$ & $0$ & $5$ \\ 
   Fig.~\ref{fig:QU_i}a  & $165$ & $90$ & $1$ & $90$ & $0$ & $0$ & $3,4.5,7$ \\ 
   Fig.~\ref{fig:QU_i}b  & $150$ & $90$ & $1$ & $90$ & $0$ & $0$ & $3,4.5,7$ \\ 
   Fig.~\ref{fig:QU_i}c  & $140$ & $90$ & $1$ & $90$ & $0$ & $0$ & $3,4.5,7$ \\ 
   Fig.~\ref{fig:QU_i}d  & $165$ & $90$ & $1$ & $0$ & $0$ & $0$ & $3,4.5,7$ \\ 
   Fig.~\ref{fig:QU_i}e  & $150$ & $90$ & $1$ & $0$ & $0$ & $0$ & $3,4.5,7$ \\ 
   Fig.~\ref{fig:QU_i}f  & $140$ & $90$ & $1$ & $0$ & $0$ & $0$ & $3,4.5,7$ \\ 
   Fig.~\ref{fig:QU_i}g  & $165$ & $90$ & $0$ & $0$ & $1$ & $0$ & $3,4.5,7$ \\ 
   Fig.~\ref{fig:QU_i}h  & $150$ & $90$ & $0$ & $0$ & $1$ & $0$ & $3,4.5,7$ \\ 
   Fig.~\ref{fig:QU_i}i  & $140$ & $90$ & $0$ & $0$ & $1$ & $0$ & $3,4.5,7$ \\ 
   Fig.~\ref{fig:QU_eta_flying_towards}(\ref{fig:QU_eta_flying_away})a  & $165$ & $90$ & $1$ & $90$ & $0$ & $0...45(-45)$ & $3,4.5,7$ \\ 
   Fig.~\ref{fig:QU_eta_flying_towards}(\ref{fig:QU_eta_flying_away})b  & $150$ & $90$ & $1$ & $90$ & $0$ & $0...45(-45)$ & $3,4.5,7$ \\ 
   Fig.~\ref{fig:QU_eta_flying_towards}(\ref{fig:QU_eta_flying_away})c  & $140$ & $90$ & $1$ & $90$ & $0$ & $0...45(-45)$ & $3,4.5,7$ \\ 
   Fig.~\ref{fig:QU_eta_flying_towards}(\ref{fig:QU_eta_flying_away})d  & $165$ & $90$ & $1$ & $0$ & $0$ & $0...45(-45)$ & $3,4.5,7$ \\ 
   Fig.~\ref{fig:QU_eta_flying_towards}(\ref{fig:QU_eta_flying_away})e  & $150$ & $90$ & $1$ & $0$ & $0$ & $0...45(-45)$ & $3,4.5,7$ \\ 
   Fig.~\ref{fig:QU_eta_flying_towards}(\ref{fig:QU_eta_flying_away})f  & $140$ & $90$ & $1$ & $0$ & $0$ & $0...45(-45)$ & $3,4.5,7$ \\ 
   Fig.~\ref{fig:QU_eta_flying_towards}(\ref{fig:QU_eta_flying_away})g  & $165$ & $90$ & $0$ & $0$ & $1$ & $0...45(-45)$ & $3,4.5,7$ \\ 
   Fig.~\ref{fig:QU_eta_flying_towards}(\ref{fig:QU_eta_flying_away})h  & $150$ & $90$ & $0$ & $0$ & $1$ & $0...45(-45)$ & $3,4.5,7$ \\ 
   Fig.~\ref{fig:QU_eta_flying_towards}(\ref{fig:QU_eta_flying_away})i  & $140$ & $90$ & $0$ & $0$ & $1$ & $0...45(-45)$ & $3,4.5,7$ \\ 
   Fig.~\ref{fig:UQ_disc_segm}a,d  & $165$ & $90$ & $0$ & $0$ & $1$ & $0$ & $3,4.5,7$ \\ 
   Fig.~\ref{fig:UQ_disc_segm}b,e  & $150$ & $90$ & $0$ & $0$ & $1$ & $0$ & $3,4.5,7$ \\ 
   Fig.~\ref{fig:UQ_disc_segm}c,f  & $130$ & $90$ & $0$ & $0$ & $1$ & $0$ & $3,4.5,7$ \\ 
   Fig.~\ref{fig:UQ_precessing_ring}  & $0...90$ & $90$ & $0$ & $0$ & $1$ & $0$ & $3,4.5,7$ \\ 
   Fig.~\ref{fig:PD_i}  & $0...90$ & $90$ & $1$ & $90$ & $0$ & $0$ & $3,7,10^4$ \\ 
\hline                                   
\end{tabular}
\end{table*}

\subsection{Rotation of polarization angle}
\label{sect:pa}

The method described above allows to compute the observed Stokes parameters once the total rotation of the polarization plane along the photon trajectory ($\chi^{\rm tot}$) is known. 
The PA seen by the observer (with respect to the direction of the disk axis) is computed as a sum of the PA of emitted photon and its total rotation:
\begin{equation}
  \chi = \chi_0 + \chi^{\rm tot},
\end{equation}
where $\chi_0=90\degr$ for synchrotron emission.
Under the assumption of photon trajectories lying in plane (corresponding to Schwarzschild metric), the total rotation can be presented as the sum of rotations caused by the light bending and relativistic aberration effects \citep[the aberration effects need to be computed in the curved space, i.e. accounting for the the deflection of the photon path, which is different from the case of the flat space][]{Pineault1977pol_schw,Loktev2022}.
We follow the methodology described in \citet{Loktev2022,Loktev2024} where the rotation angles are computed relative to the normal of the disk (in our case, equatorial plane).
As the PA of synchrotron radiation is related to the magnetic field direction, we first need to link this direction to the disk normal via the rotation angle $\chi^B$.
Then, the rotation of polarization plane resulting from light bending is described by the angle $\chi^{\rm{GR}}$ and the special relativity (aberration) effects are accounted for by the angle $\chi^{\rm{SR}}$.
This leads to the total rotation angle through the sum
\begin{equation}
    \chi^{\rm{tot}} =\chi^B + \chi^{\rm GR} + \chi^{\rm SR}.
\end{equation}

The explicit expression for the rotation caused by the effects of special relativity is given in Appendix~\ref{sect:appendixA}:
\begin{equation}\label{eq:chiSR}
    \tan\chi^{\rm SR} = - \beta \frac{\cos\alpha\cos\zeta}{\sin^2\zeta - \beta \cos\xi}.
\end{equation}
The form of this expression is the same as in eq.~(31) of \citet{Loktev2022}, however, the terms are not the same.
Namely, in this work we explicitly consider the cases of a non-zero accretion velocity ($\sigma \neq 90\degr$ or $270\degr$) and out-of-equatorial-plane matter location ($\eta \neq 0$), hence the expressions for the terms $\cos\zeta$, $\sin\zeta$ and $\cos\xi$ are generally different as compared to the ones given in the former work (see Eqs.~\ref{eq:coszeta} and \ref{eq:cosxi}).
In the case of the flat (Minkowski) space, the expression reduces to
\begin{equation}\label{eq:chiSRflat}
    \tan\chi^{\rm SR}_{\rm flat} = - \beta \tan i 
    \frac{\sin i \cos\varphi \cos\eta + \cos i \sin\eta}{\sin i - \beta \cos(\varphi+\sigma)}.
\end{equation}
For $\eta=0$ and $\sigma=90\degr$ it reduces to eq.~(32) of \citet{Loktev2022}.

The expression for $\chi^{\rm GR}$ becomes lengthy once we consider photons originating from the plane that is not equatorial.
In coordinate form it reads as (see Appendix~\ref{sect:appendixA})
\begin{equation}
    \tan\chi^{\rm GR} = \cos\eta \frac{\Tilde{a}_1 \sin\psi - \Tilde{a}_2 \sin i \sin \varphi}{\Tilde{a}_1\Tilde{a}_2 - \cos^2\eta \sin^2 i \sin^2\varphi \sin\alpha},
\end{equation}
where 
\begin{eqnarray}
    \Tilde{a}_1 &=& \sin\eta + \cot\psi \sin i \cos i \sin\alpha \sin\varphi, \\
    \Tilde{a}_2 &=& \sin\alpha \left[ \cos\alpha \cos i - \sin\eta \cos(\psi-\alpha)\right].
\end{eqnarray}
For $\eta=0$ this expression reduces to eq.~(29) of \citet{Loktev2022}.
 
The rotation angle transforming the polarization direction from the local magnetic field direction relative to the disk normal can be expressed as (see Appendix~\ref{sect:appendixA})
\begin{equation}\label{eq:PAmagnvector}
     \chi^{\rm{B}}=\arctan\left( \frac{\unit{k}^{\prime}_0\cdot(\unit{n}\times\unit{B'})}{\unit{n}\cdot\unit{B'}-(\unit{k}^{\prime}_0\cdot\unit{n})(\unit{k}^{\prime}_0\cdot\unit{B'})} \right).
\end{equation}
The expression for $\chi^B$ takes into account the both light bending effects ($\psi\neq\alpha$) and the Lorentz transformation of the magnetic field (primed quantities).
This angle is related to both general and special relativity effects, and depends on $\alpha$ and $\delta$.
Explicit expression for the angle $\chi^B$ reads as
\begin{equation}\label{eq:chiB}
    \tan\chi^{B} = \frac{\left[ \Tilde{b} \cos\phi' + \gamma\sin\phi' (\cos\xi - \beta)\right]}
    {\delta \cos\zeta \left[ \gamma \cos\phi'(\cos\xi - \beta) + \Tilde{b}\sin\phi'\right]
    +\frac{b_{z}}{b_{\rm eq}} \left[ \delta\cos^2\zeta - \frac{1}{\delta})\right]},
\end{equation}
where $\Tilde{b}$ is taken from Eq.~\eqref{eq:tildeb}.

Expression for $\chi^{B}$ for purely circular rotation ($\sigma=90\degr$) in the equatorial plane ($\eta=0$), far from the compact object ($\gamma\to1$, $\beta\to 0$) and small, yet non-zero inclinations ($i\to0$) reduces to
\begin{eqnarray}
    &\tan{\chi^{B}}& \approx \tan\varphi, \,\,\, {\rm if}\; \phi'=90\degr,\\  \nonumber
    &\tan{\chi^{B}}& \approx - \cot{\varphi}, \,\,\, {\rm if}\; \phi'=0.
\end{eqnarray}
Hence, we obtain
\begin{eqnarray}\label{eq:chiB_far}
    & \chi^{B} & \approx  \varphi \;\;\; {\rm for~radial~field},  \\ \nonumber
    & \chi^{B} & \approx  \varphi-90\degr \;\;\; {\rm for~toroidal~field}.
\end{eqnarray}
The observed PA is then $\chi=\varphi+90\degr$ for radial and $\chi=\varphi$ for toroidal fields, and in both cases the result matches the intuitive expectations.

\begin{figure*}
\centering
         \includegraphics[width=\linewidth]{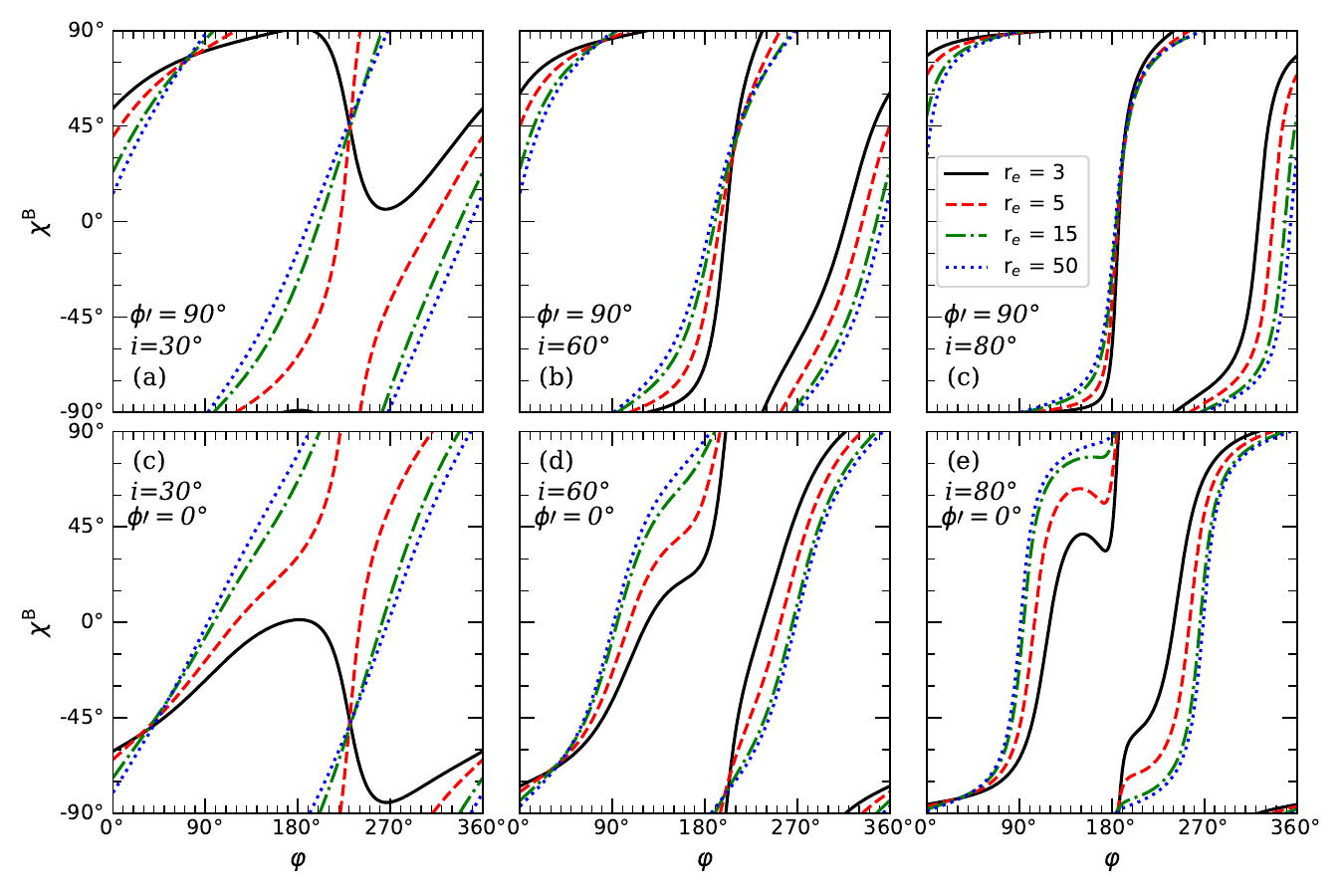}
   \caption{
   Rotation of polarization angle due to magnetic field orientation, as a function of azimuth (in disk coordinates) for different equatorial radii (measured in units of Schwarzschild radii): $r_{\rm e}=3$ (black solid), $5$ (red dashed), $15$ (green dot-dashed) and $50$ (blue dotted). The matter is assumed to undergo circular, counter-clockwise rotation ($\sigma=90\degr$) with velocity following Luminet's law. Upper panels correspond to purely radial co-moving magnetic field ($\phi'=90\degr$) and lower panels correspond to purely toroidal field ($\phi'=0$). Columns (left to right) correspond to the observer inclination $i=30\degr$, $60\degr$ and $80\degr$. 
   }\label{fig:PAinitial}%
    \end{figure*}

\section{Results}
\label{sect:results}

The developed formalism allows to analytically compute the polarimetric signatures of matter close to the event horizon.
The power of such formalism is that we can separately understand each effect contributing to the total modification of polarization, hence develop an intuition on the range of the model parameters for the given observed set of data.
Furthermore, using the developed formalism, we can compare the predictions of the polarization rotation to the synchrotron source in an orbit around the BH.

Many parameters affect the polarimetric properties of the synchrotron source: distance to the BH $r$, $B$-field topology, observer inclination $i$, location of the source with respect to the disk plane $\eta$ and the direction of velocity vector as determined by $\sigma$.
Generally, with increasing distance to the BH $r$ all relativistic effects fade and only the topology of magnetic field affects the PA for different parts of the disk.
Other effects are discussed below in detail.
To produce the figures, we consider point source at given coordinates, unless stated otherwise.

\subsection{Rotation of PA due to relativistic effects}
\label{sect:relativistic_effects}

To disentangle the rotation of the polarization plane caused by relativistic effects from the effect of the B-field orientation changes, we first consider the case of a vertical field, $B=(0,0,B_z)$. 
In the absence of the relativistic effects, this case is trivial: the PA is constant and equal to $90\degr$ (for all inclinations except for the degenerate $i=0$).
In the absence of relativistic effects, that can alter the direction of the magnetic field, the angle is the same across the entire disk.
We have verified that our calculations reproduce this value with high accuracy.

We then include the special and general relativistic effects and show the contours of constant PA as a function of the disk coordinates $(r,\varphi)$ in Fig.~\ref{fig:chiGRSR_contours}.
We assume circular counterclockwise rotation $\sigma=90\degr$ and consider different inclinations $i=30\degr$, $60\degr$ and $80\degr$ and various elevations above the disk plane $\eta=0,\pm20\degr$.
We note that the cases $\eta\neq0$ correspond to conical surfaces above/below the disk plane, but very similar contours can be obtained by fixing the height $h=3\tan{\eta}$ for all radii.
All parameters are listed in Table~\ref{table:polar}.
The azimuthal angle in the disk grows from the direction $(X,-Y)$.
For the case of rotation in the equatorial plane ($\eta=0$) the contours of $\chi^{\rm GR}+\chi^{\rm SR}$ are essentially the same as shown in fig.~4 of \citet{Loktev2022}.
We note the presence of a critical point at $i=30\degr$ near coordinates $(-3,2)$, where all lines of constant PA rotation intersect. 
This point corresponds to the place in the disk that an observer sees face-on as a result of the light bending and SR effects, hence the PD=0 in this point, leading to a degeneracy of PA.

Movement around the origin in Fig.~\ref{fig:chiGRSR_contours} corresponds to a rotation of a point (infinitely small spot) in a disk at a certain radius.
The radius containing the critical point separates the regions where the rotation of polarization angle crosses the $\pm90\degr$ cycle point from the region where the PA does not make a cycle.
These two regions correspond to either one or two loops made by Stokes parameters per one orbital cycle of a spot, and existence of such critical point has been previously noticed in a number of works \citep[e.g.,][]{Dovciak2008,GRAVITY2018,GRAVITY2020_analytical,Gelles2021,Vos2022,Loktev2022,Vincent2023}.

\begin{figure*}
 \centering
 \includegraphics[width=0.82\linewidth]{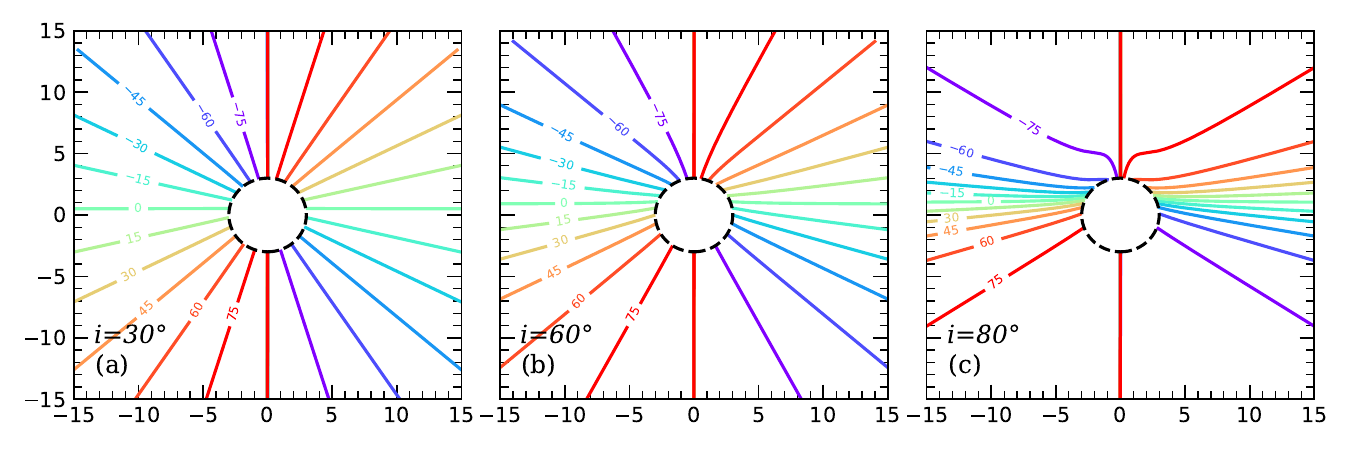}
 \caption{Contours of constant total rotation angle ($\chi^{\rm tot}$) for the case of toroidal magnetic field ($\sigma=90\degr$, $\phi'=0$) and the equatorial plane matter movement, in disk coordinates. We note the disappearance of the critical point, as compared to the case of vertical field, as the rotation is dominated by the changing magnetic field orientation ($\chi^{\rm B}$). The observed PA $\chi$ is obtained by adding $\chi_0=90\degr$.}
\label{fig:critical_Bphi}
\end{figure*}

The location of a critical point have been discussed in detail in \citet{Vincent2023} under the assumption of Minkowski space.
However, its appearance is not entirely caused by aberration effects, and there are essential differences with respect to the contour morphology in Schwarzschild metric.
First, the SR rotation itself is different in Minkowski and Schwarzschild metric  (compare Eq.~\ref{eq:chiSR} and \ref{eq:chiSRflat}), as the $\unit{k}_0$ and $\unit{o}$ directions are not aligned.
Inclusion of the light bending within the calculation of the PA rotation caused by aberration mainly leads to a strong up-down contours asymmetry (around the horizontal line $y=0$ in Fig.~\ref{fig:chiGRSR_contours}) and a shift of the position of the critical point from the azimuth $\varphi=270\degr$ for flat space towards $\varphi\approx230\degr$ for Schwarzschild space.

Interestingly, pure aberration effects in Schwarzschild metric may lead to a formation of two critical points at different radii \citep[fig.~6 in][]{Loktev2022}, however, the additional rotation due to light bending effects alter their location and may lead to a shift of one of them below ISCO.
This effect is seen in Fig.~\ref{fig:chiGRSR_contours}: only one critical point is observed in the panels (a), (d) and (g), corresponding to $i=30\degr$.
At higher inclinations the critical point disappears because the joint action of aberration and light bending effects are not strong enough to direct the photons escaping vertically from the disk into the observer's direction and hence the PA rotation is always smaller than one complete cycle.

The radius and azimuthal coordinate of the critical point, along with the contours morphology is also altered if the matter is located above or below the equatorial plane (middle and lower panels in Fig.~\ref{fig:chiGRSR_contours}).
Since in our assumptions we considered the same speed of matter at different elevations, the effect purely scales with the magnitude of light bending.

\begin{figure*}
\centering
\includegraphics[width=\linewidth]{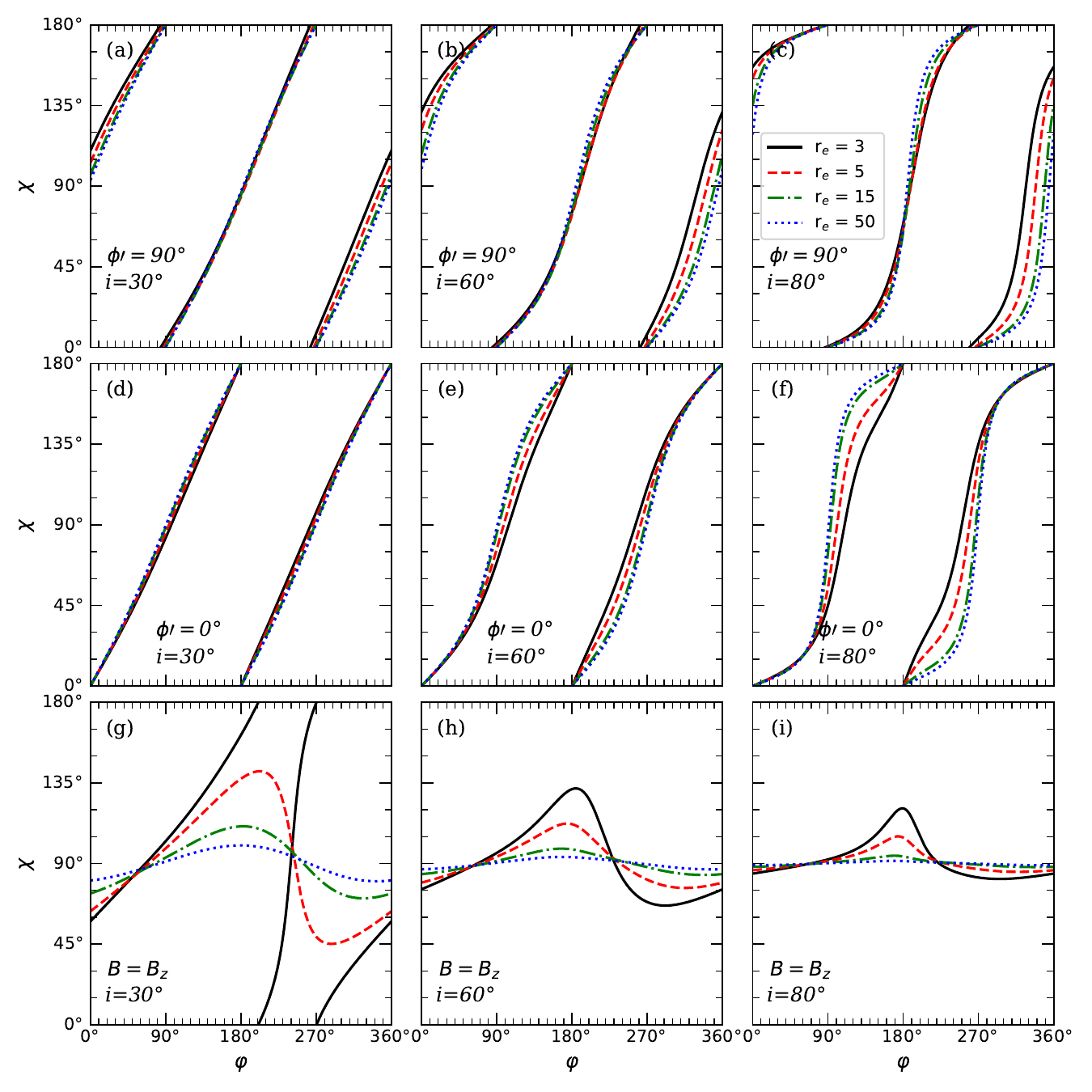}
\caption{The PA seen by the observer ($\chi$) obtained for the angle $\eta=20\degr$, for different radii, observer inclinations and magnetic fields, as a function of azimuth in the disk. 
Upper panels show the case of radial field, middle panels correspond to purely toroidal field and lower panels correspond to the vertical field.
Left to right columns correspond to an increasing observer inclination: $i=30\degr$, $60\degr$ and $80\degr$.
The matter is assumed to rotate counter-clockwise ($\sigma=90\degr$), with velocity following Luminet's law.}
      \label{fig:chitot_eta20}
\end{figure*}

\begin{figure*}
\centering
\includegraphics[width=\linewidth]{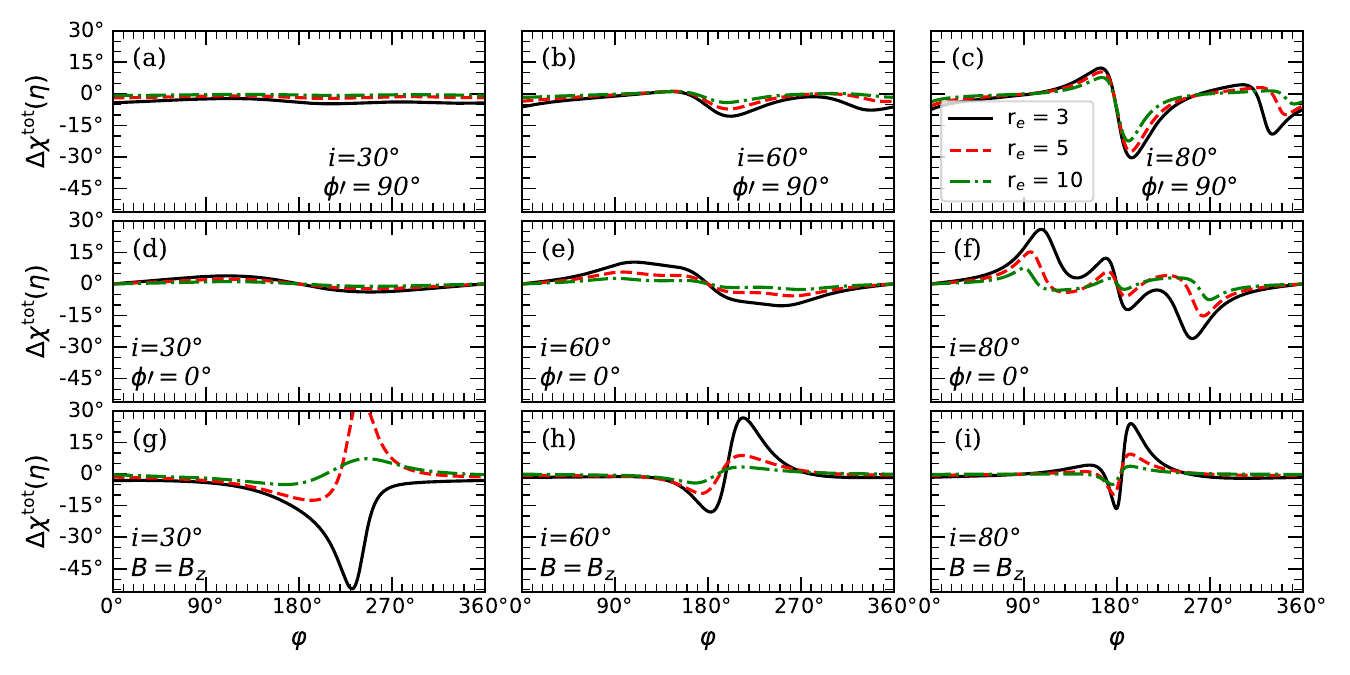}
\caption{Difference between the PA obtained for the angle $\eta=20\degr$ and $\eta=0$ ($\chi^{\rm tot}(\eta=20\degr)-\chi^{\rm tot}(\eta=0)$), for different radii, observer inclinations and magnetic fields, as a function of azimuth in the disk coordinates. 
Upper panels show the case of radial field, middle panels correspond to purely toroidal field and lower panels correspond to the vertical field.
Left to right columns correspond to an increasing observer inclination: $i=30\degr$, $60\degr$ and $80\degr$.
The matter is assumed to rotate counter-clockwise ($\sigma=90$), with velocity following Luminet's law.}
          \label{fig:diff_chitot_eta20}
\end{figure*}

\begin{figure*}
\includegraphics[width=\linewidth]{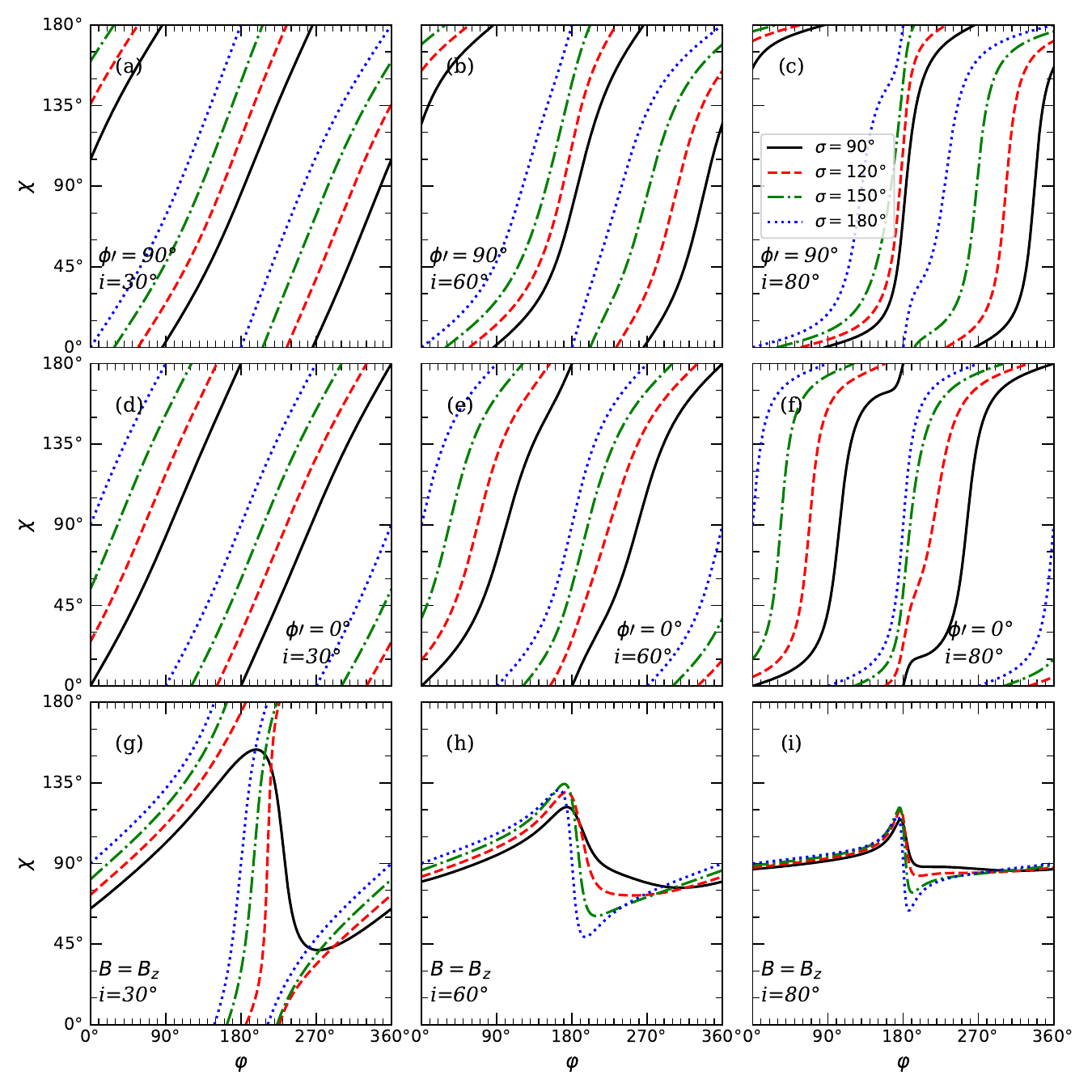}
\caption{Effects of the non-circular motion ($\sigma$) on the observed PA ($\chi^{\rm tot}$) for different inclinations (left to right: $i=30\degr$, $60\degr$ and $80\degr$) and magnetic fields (upper panels: radial field, middle panels: toroidal field, lower panels: vertical field), as a function of the azimuth in the disk coordinates. 
Different direction of motions correspond to $\sigma=90\degr$ (black solid lines), $110\degr$ (red dashed), $150\degr$ (green dot-dashed) and $180\degr$ (blue dotted).
The emitting matter is assumed to be located in the equatorial plane ($\eta=0$) at distance $r=5$, with dimensionless velocity obeying the Luminet's law.}
 \label{fig:chi_sigma}
\end{figure*}

\subsection{Effects of magnetic field orientation}
\label{sect:magnetic_field-effects}

Additional rotation of PA can be gained because of the varying orientation of magnetic field with azimuth, e.g. for the cases of purely toroidal and radial magnetic fields.
In Fig.~\ref{fig:PAinitial} we show the results of calculations of $\chi^{B}$, as a function of the disk azimuth $\varphi$, for these two cases and three different inclinations of the observer: $i=30\degr$, $60\degr$ and $80\degr$.
Circular counterclockwise ($\sigma=90\degr$) rotation of matter at the equatorial plane of the disk ($\eta=0$) is assumed.
All parameters are listed in Table~\ref{table:polar}.
Different lines correspond to matter located at different equatorial radii: $r_{\rm e}=3$ (black solid), $5$ (red dashed), $15$ (green dot-dashed) and $50$ (blue dotted).

For radii $r_{\rm e}\gtrsim 50$ the angle makes two full loops per azimuthal cycle for any inclination.
For the simplest case of the disk at relatively low inclination, $i=30\degr$, the rotation $\chi^{\rm B}$ increases with azimuth nearly linearly.
This replicates the behavior for small inclinations and large distances to the compact object described in Eq.~\eqref{eq:chiB_far}.

For the radii $r_{\rm e}\lesssim 5$ there are noticeable deviations from the behavior at large radii thanks to the pronounced effects of relativistic aberration, modified by the light bending.
Difference is particularly high at azimuth $\varphi\sim270\degr$, i.e. for the approaching side of the disk. 
For $i=30\degr$ the changes are severe, as rotation reduces to one loop over the azimuthal cycle, hence the whole topology of rotation changes. 
Behavior of PA with the azimuth is now different for the radial and toroidal fields, making it potentially possible to distinguish these fields via orbital (azimuthal) phase-resolved polarimetric data.

For both toroidal and radial fields the total PA rotation ($\chi^{\rm tot}$) is dominated by the rotation caused by the different magnetic field orientation ($\chi^{B}$), rather than by the rotation caused by relativistic effects ($\chi^{\rm GR}+\chi^{\rm SR}$), and the contours of constant rotation do not show contain the critical point (examples of contours for the case of toroidal field and different $i$ are shown in Fig~\ref{fig:critical_Bphi}).
However, the lines of PA rotation have nearly the same morphology for pure radial and toroidal fields, albeit shifted in azimuth.
Hence, their linear combination may lead to a substantial reduction of the PA rotation with azimuth, essentially mimicking the vertical field and may replicate the behavior near the critical point (Fig.~\ref{fig:chiGRSR_contours}).
We find that the critical point appears e.g. for the combination of radial and toroidal field components ($\phi'=45\degr$) at high inclinations ($i=80\degr$) and matter below the equatorial plane ($\eta=-20\degr$).

\subsection{Effects of non-equatorial motion}

The GR and SR effects on rotation of polarization plane have been studied, using the aforementioned formalism, for the case of a razor-thin matter in an equatorial plane of BHs \citep{Loktev2022,Loktev2024}.
Emitting matter may however be elevated above the equatorial plane in the case of geometrically thick disks or if emission arises at the interface between the inflow and outflow \citep[edge-brightened jet found, e.g. in high-resolution images of Cen~A,][]{Janssen2021}.
Our formalism enables direct comparison between the polarization rotation for the matter rotating in the plane containing the BH itself (equatorial plane, $\eta=0$) and that of the matter elevated by the height $h$ (set by $\eta\neq0$) above/below this plane.

In Fig.~\ref{fig:chitot_eta20} we show the resulting observed PA ($\chi$) for the specific case of $\eta=20\degr$. 
The topology of lines is the same as for the case $\eta=0$, but the absolute values differ.
For the cases of equatorial field (two upper rows of panels), the PA rotation is dominated by the $\chi^{B}$ term and GR/SR effects are less pronounced, as lines corresponding to different radii nearly overlay, especially for smaller inclinations.
Fig~\ref{fig:chiGRSR_contours} (d-f) shows this effect in more detail. 
The contours of the same $\chi^{\rm GR}+\chi^{\rm SR}$ are generally closer to the event horizon for $\eta=20\degr$, as compared to $\eta=0$. 
This means that for the same distance from the BH, the polarization plane rotation is more severe for the case of equatorial rotation than for the matter located between this plane and the observer.
The main reason for this is that for higher $\eta$, smaller part of photon path lies close to the strong gravitational field of the BH, hence bending effects are less pronounced.
Likewise, if the matter is located below the equatorial plane, i.e. behind the BH, relativistic effects become most pronounced and contours of higher PA rotation appear (Fig.~\ref{fig:chiGRSR_contours}g-i).

To quantify the effects of matter elevation at different radii, in Fig.~\ref{fig:diff_chitot_eta20} we show the difference between the total rotation ($\chi^{\rm tot}$) for the case of $\eta=20\degr$ with respect to the case $\eta=0$ for various inclinations and magnetic field configurations, observer inclinations and radii to the BH.
For large distances $r$ the effect of elevated matter is minor, as expected. 
However, for the distances $r\lesssim10$ the difference can reach $30\degr$, becoming comparable to the total rotation, hence cannot be neglected.
The difference can reach $90\degr$ for certain parameters and azimuths.
It is generally asymmetric with respect to $\varphi$ and has complex dependence on the magnetic field orientation, making it difficult to {\it a-priori} predict the effect of elevated matter onto the PA rotation.
However, it is generally clear that the effect is highest at azimuths $180\lesssim\varphi\lesssim270$, where the GR and SR effects are most pronounced.
Since these effects strongly depend on the distance to the compact object and matter velocity, any changes to the distance caused by non-zero elevation result in substantial changes of total PA rotation.

\subsection{Effects of non-circular motion}

Accretion flows around black holes have a non-zero radial velocity component -- the accretion speed.
In our formalism, this is parameterized by the angle $\sigma$, which may potentially affect the resulting rotation of polarization, as highest aberration may be achieved at a different angle.
The trivial case of changing direction is the clockwise rotation, $\sigma=270\degr$, which can be visualized by simply changing the azimuthal angle $\varphi_{\rm cl-w}=360\degr-\varphi$.
Another important example includes spherical (Bondi-Hoyle) accretion, $\sigma=180\degr$.
We consider the effects of non-circular motions of matter for different magnetic field geometry and varying inclinations in Fig~\ref{fig:chi_sigma}.

For equatorial field cases (panels a--f), the field direction is always fixed either parallel or orthogonal to the matter motion, hence when the velocity direction changes, the field direction follows, introducing constant shifts to the lines.
Apart from the constant shifts, there are minor deviations of the PA dependence on the azimuthal angle owing to the different combined SR and GR effects.
Again, these are mostly pronounced for the locations of matter behind the BH ($\varphi\sim180\degr$).

In the case of the vertical field, the changes are generally likewise minor (Fig.~\ref{fig:chi_sigma}h,i).
The highest difference with respect to the default counter-clockwise rotation occurs for $i=30\degr$ (Fig.~\ref{fig:chi_sigma}g), close to the location of the critical point.
By changing the velocity vector, it is possible to shift this point, changing the character of the PA rotation from the limited-range (black solid line in Fig.~\ref{fig:chi_sigma}g) to two complete loops per orbit (red dashed, green dot-dashed and blue dotted).
Hence, the $\sigma$ parameter may be degenerate with inclination.

\begin{figure*}
 \centering
 \includegraphics[width=0.8\linewidth]{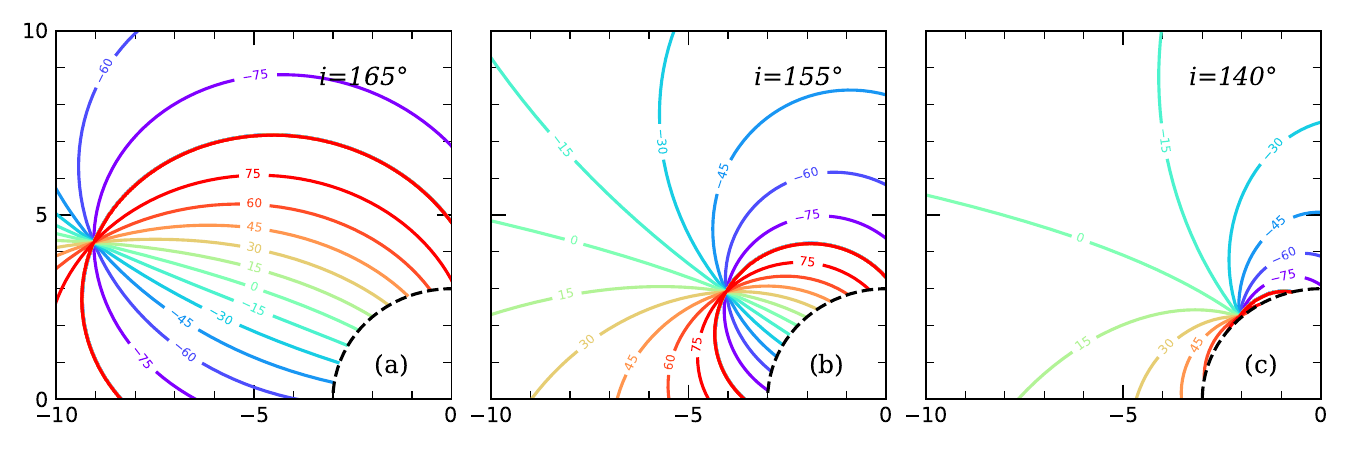}
 \caption{Contours of constant GR and SR rotation angles ($\chi^{\rm GR}+\chi^{\rm SR}$), in disk coordinates. Position of the critical point as a function of inclination for rotation caused solely by GR and SR effects (equivalent to $B=B_z$). The disk rotates counterclockwise ($\sigma=90\degr$) in the equatorial plane, but at inclinations $i=165\degr$ (a), $155\degr$ (b) and $140\degr$ (c) it is seen as rotating in the clockwise direction.}
\label{fig:critical_Bz}
\end{figure*}

\begin{figure*}
\centering
\includegraphics[width=0.75\linewidth]{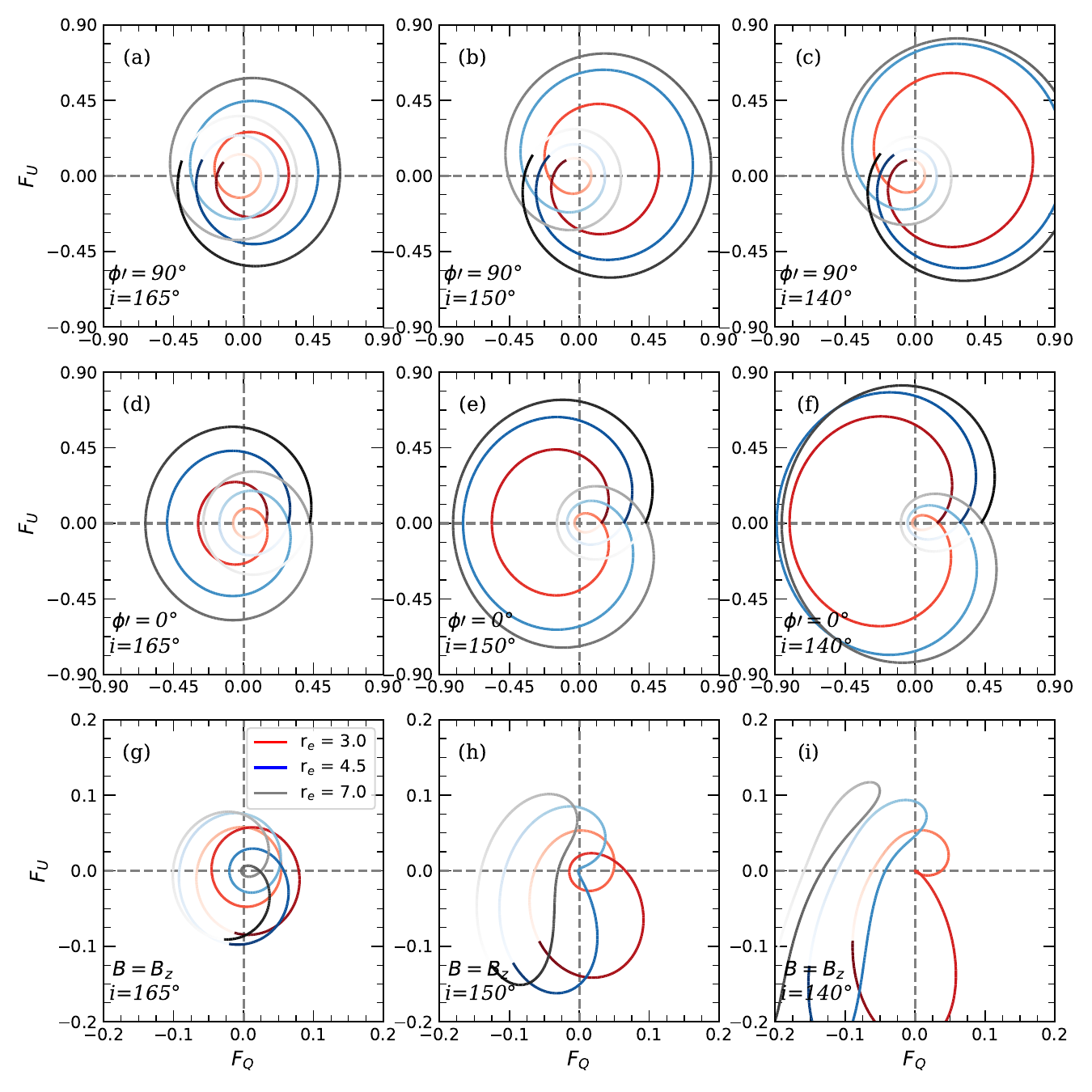}
\caption{Flux Stokes parameters $F_Q$ and $F_U$ obtained using $\eta=0$, for different radii, $B$ and $i$. Keplerian velocity and $\sigma=90\degr$ are assumed (note that for the considered inclinations the observed matter movement is clockwise on the sky). Increasing saturation of color corresponds to progression of the spot along the azimuth. The time whet the line intersects zero PD corresponds to the passage through the critical point. Note different scales in figures.}
 \label{fig:QU_i}
\end{figure*}

\section{Applications}
\label{sect:applications}

In this section we compare the topology of static and dynamic signatures of BHs.
The developed formalism can be applied to reproduce the horizon-scale images and polarimetric properties of M87* and Sgr~A* \citep{EHT2019I,EHT2021VIII,EHT2022I_Sgr,EHT2024VII_Sgr}, as well as dynamic polarimetric signatures of Sgr~A* \citep{Wielgus2022,GRAVITY2023}.
We start with the expected polarimetric signatures of the orbiting compact and extended spots.
We then discuss polarimetric signatures of the narrow disk ring.
After that, we show examples of computed images of the geometrically thick accretion disk.

\begin{figure*}
\centering
\includegraphics[width=0.8\linewidth]{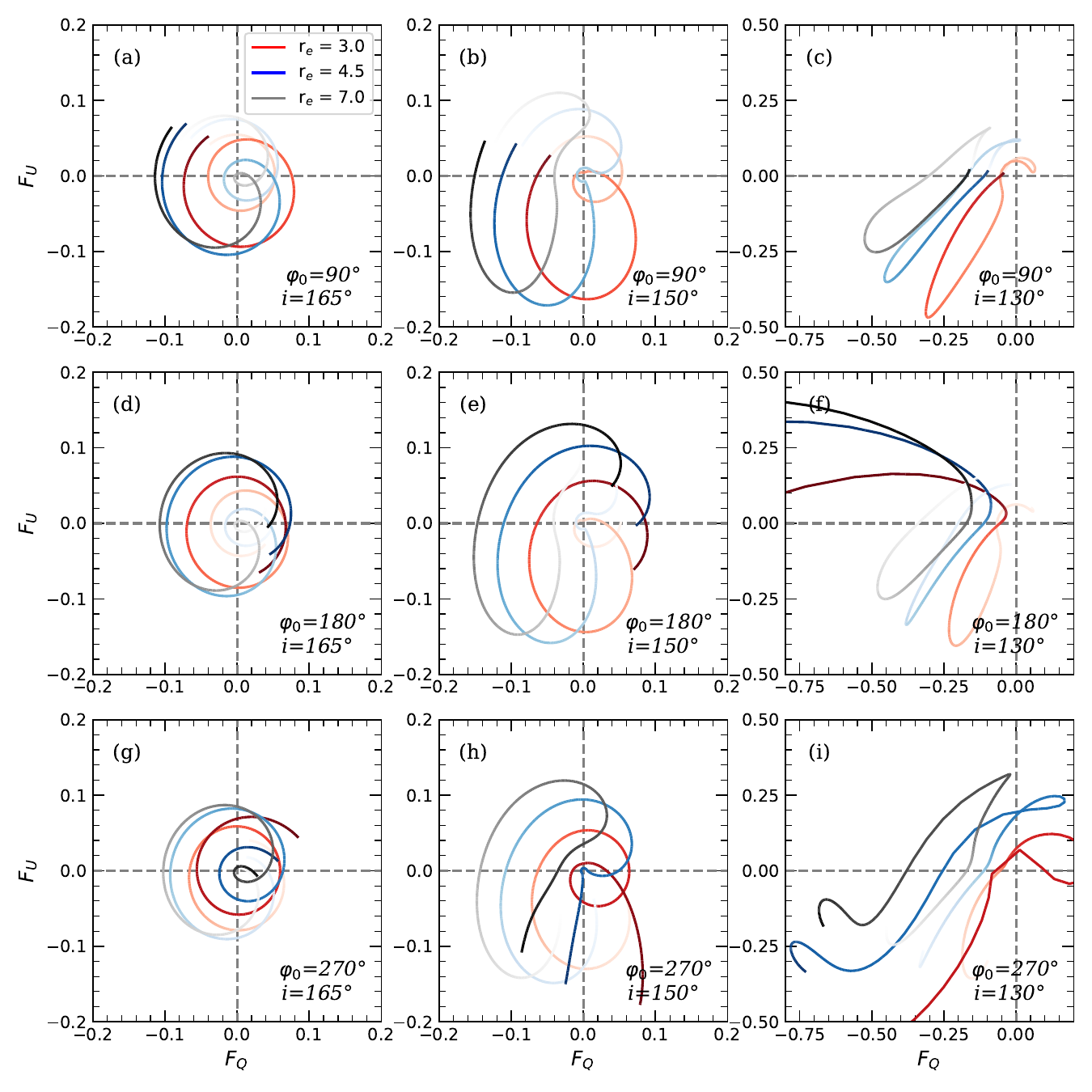}
\caption{Stokes parameters $U$ and $Q$ for a flying matter towards the observer ($\eta_{\rm max}=45\degr$), in the direction orthogonal to rotation plane, at different launching azimuthal angles: $\varphi_0=90\degr$ (upper panels), $180\degr$ (middle panels) and $270\degr$ (lower panels). Note different scales in panels.}
          \label{fig:QU_eta_flying_towards}
\end{figure*}

\begin{figure*}
\centering
\includegraphics[width=0.8\linewidth]{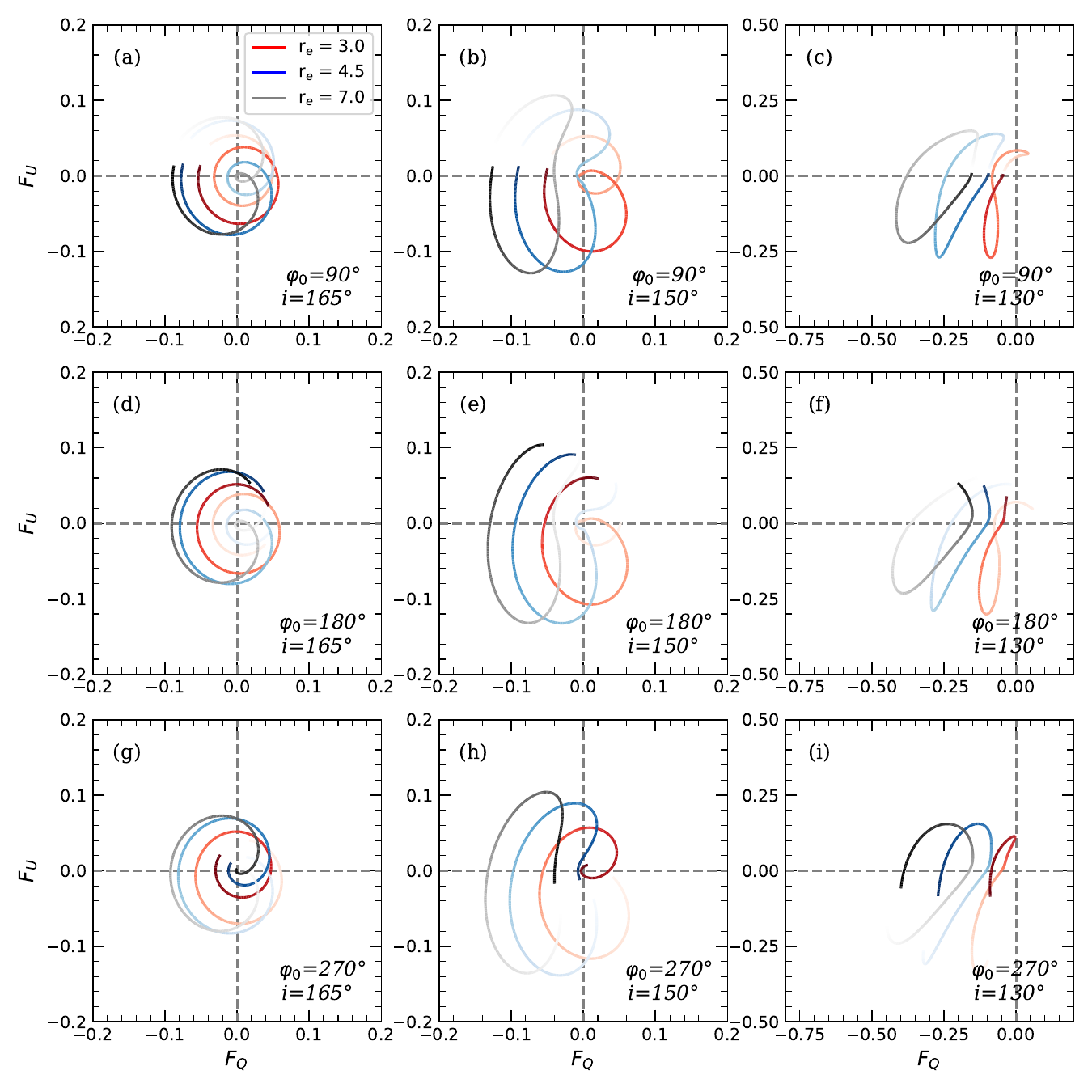}
\caption{Same as in Fig.~\ref{fig:QU_eta_flying_towards}, but for the matter flying away from the observer ($\eta_{\rm max}=-45\degr$) in the direction orthogonal to the rotation plane. Note different scales in panels.}
          \label{fig:QU_eta_flying_away}
\end{figure*}

\begin{figure*}
\centering
\includegraphics[width=0.8\linewidth]{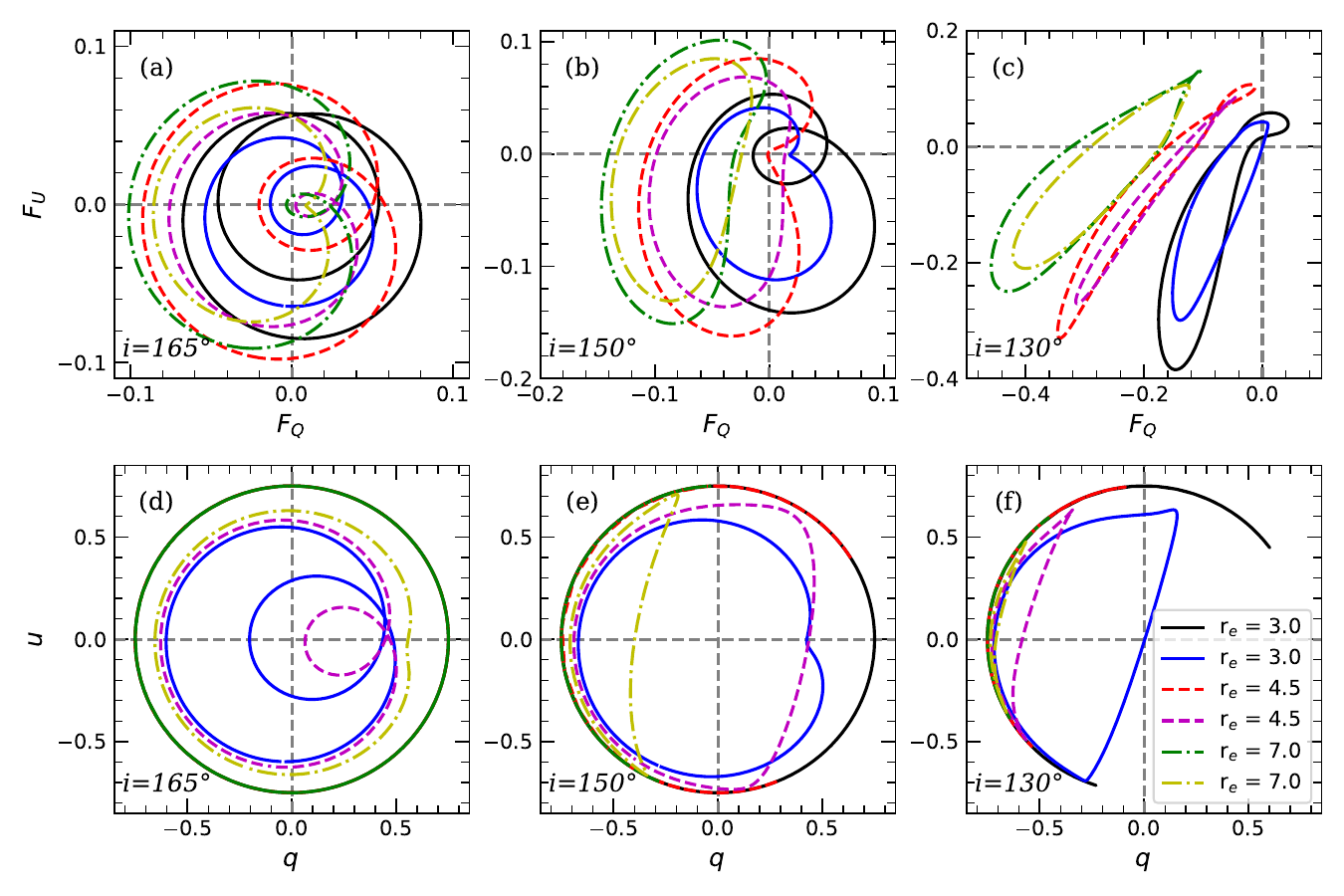}
\caption{Stokes parameters of the infinitely compact spots and disk segments. Averaging was made over $\Delta\varphi=45\degr$ (i.e., quarter of the circle). Upper and lower panels correspond to the absolute and relative Stokes parameters, respectively. Black, red and green lines denote calculations of the infinitely small spot, while blue, violet and yellow lines correspond to the parameters averaged over the segment. Note that several lines overlap in the lower panels.
}\label{fig:UQ_disc_segm}
\end{figure*}

\subsection{Hot spot rotation}
\label{sect:hot_spot_rotation}

Quiescent level of Sgr~A* accretion is known to be interrupted by the bright flares, when the observed radio (sub-mm) and NIR fluxes can suddenly increase by 1--2 orders of magnitude \citep[e.g.][]{Baganoff2001,Genzel2003}.
The characteristic timescales of the flares are roughly consistent with Keplerian time close to the event horizon, and the hot spot rotation scenario was proposed to explain the flares \citep{Broderick2006,Trippe2007}.
Alternatively, a pattern movement could be responsible for the production of similar characteristics \citep{Matsumoto2020,Aimar2023}.
Recently-observed movement of the brightness centroid and loops of polarized fluxes have given additional support to these scenarios \citep{GRAVITY2018,Wielgus2022,GRAVITY2023}.
Perhaps, the most intriguing property found in those data is the observed one polarimetric loop in Stokes parameters per one orbital period, with zero-polarization point (where absolute Stokes parameters $Q=U=0$) being located within this loop (in the NIR, where interstellar polarization is expected to be low).
Moreover, the PA was found to change at a nearly constant rate as a function of time, presumably corresponding to the azimuthal angle in the accretion disk \citep{GRAVITY2023}.

One complete rotation of PA per orbit is not expected to appear for a large fraction of parameter space and is tightly linked to the location of the critical point (where the observed polarization turns zero).
If the orbit does not intersect the location of a critical point $(r_{\rm cr}, \varphi_{\rm cr})$, either two complete loops of Stokes parameters (for orbits within $r_{\rm cr}$) or one loop that does not enclose the origin $Q=U=0$ (orbits beyond the radius of the zero-polarization point) are generally expected (see Sect.~\ref{sect:relativistic_effects}).
Appearance of the critical point is also tightly related to the magnetic field direction (Sect.~\ref{sect:magnetic_field-effects}): if the spot is moving in the equatorial plane, the critical point is only observed for the case of vertical field $B=B_z$.
In order to reproduce the observed topology, the radius and inclination of the hot spot should well coincide with the location of the critical point, so that the second loop, that occupies a small fraction of the orbital motion, was small or completely absent, and the rotation around the secondary loop was fast enough, so that it did not affect the constant rate of PA changes.
In other words, for the data that are limited in time and angular resolution, it should not be feasible to resolve the movement along the second polarimetric loop.

The limited range of radii and inclinations for which the critical point exists, as well as the strong dependence on the magnetic field topology then translates to highly constraining estimates on these parameters that are coming from polarimetry \citep[e.g.,][]{GRAVITY2018,GRAVITY2023}.
In Fig.~\ref{fig:critical_Bz} we show the contours of PA rotation for the case of vertical field, with inclinations over $90\degr$ that are believed to be relevant to the accreting matter around Sgr~A*.
For the inclinations outside $140\degr<i<165\degr$, the location of the critical point is either outside of the $\sim10\rs$, corresponding to scales that the instruments are sensitive to, or within the ISCO.

In Fig.~\ref{fig:QU_i} we show the patterns traced by the spot rotating in the equatorial plane.
Increasing time(/azimuth) is shown as saturating color: pale part of the curves correspond to the beginning of the rotation from $\varphi=0$.
To produce this figure, we use the latter representation of (Stokes) fluxes in Eq.~\eqref{eq:flux}, and keep $\rmd S'=const$ throughout the rotation. 
While the Eq.~\eqref{eq:flux} was derived assuming the spot emits through the upper surface of the disk, the same expression is also valid for the synchrotron emission from a spherical, optically thin spot of radius $h_0$, as in this case the factors $\cos\zeta$ entering the area $\rmd S'$ and the photon path $l_{\rm ph}'$ are omitted.

The loop topology becomes altered if the matter gradually rises above/below the equatorial plane.
This scenario is expected, e.g. if the spots correspond to ejecting plasmoids \citep{Aimar2023}.
We illustrate this case by assuming a gradual increase of matter elevation above the orbital plane, keeping the equatorial radius of matter the same (helical motion).
The results of calculations for the case of the vertical magnetic field and different inclinations are shown in Figs.~\ref{fig:QU_eta_flying_towards} and \ref{fig:QU_eta_flying_away}.
The topology of the loops strongly depend on the azimuthal angle $\varphi_0$ at which the spot starts rising, as evident from the comparison of the top, middle and bottom panels.

Next, we consider the behavior of relative Stokes parameters in the regime of strong gravity and fast motions.
The PD is Lorentz invariant, hence compact spots located in any part of the disk are expected to have the PD that corresponds to the PD of intrinsic emission from the corresponding disk segment viewed at an angle $\zeta'$.
Angular dependence of intrinsic PD, e.g. for the case of the optically thick disk atmospheres, leads to a modulation of the PD with orbital phase, since the viewing angle $\zeta'$, for a given observer's inclination and radius, changes with azimuth \citep[e.g., fig. 7 of][]{Loktev2022}.
However, for the case of the synchrotron emission the PD does not depend on the viewing angle, but only on the index of the power-law electrons \citep{Ginzburg1965}.
Hence, in the simple scenario we do not expect changes of the PD with orbital phase (azimuth), that translates to the topology of 2 complete cycles or a part of the cycle pattern in the $qu$ plane.
In the lower panels of Fig.~\ref{fig:UQ_disc_segm} we show the direct calculation of the relative Stokes parameters for the case of an infinitely small spots at different radii and inclinations (black, red and green curves) that confirm the two-cycle or part-cycle pattern.
A more complex pattern may be obtained if the angular dependence of the PD is caused by the angular-dependent (with respect to the magnetic field orientation) electron spectral index $p$.

\subsection{Polarimetry of an extended spot and disk ring}

An important property of the observed signal is the considerably lower PD detected in the flares, $\sim25$\%, as compared to the theoretical expectations, $\sim70$\%.
The reduction of PD may be caused by the beam depolarization \citep{Narayan2021}, or if the spot is extended in azimuthal, vertical or/and radial direction \citep[see examples of simulations in][]{Yfantis2024}.
In Fig.~\ref{fig:UQ_disc_segm} we compare the absolute ($F_{Q}$, $F_{U}$) and relative ($q,u$) Stokes parameters for the infinitely compact spot (black, red and green curves) to those that are averaged in the azimuthal direction (infinitely thin disk segment), assuming the spot size of a quarter circle (blue, violet, yellow curves).
The spot is assumed to move and each point in the plot corresponds to the azimuth of the middle of the spot.
The depolarization essentially occurs because of the averaging over Stokes vectors with different angles, but a large spot size (comparable to the orbital scale) is needed.
The single loop and distorted shapes of the $qu$ parameters can be recovered.

It is noteworthy that the behavior of NIR polarization during the flares do not always follow the one-loop pattern over the $\sim70$~min timescale (thought to correspond to the orbital period of the spots).
Namely, the pattern detected in flares from Aug 18 2019 and May 19 2022 \citep[fig.~1 of][]{GRAVITY2023} resembles a line, rather than a loop.
Furthermore, simultaneous high-precision astrometric measurements available for the flare on May 19 2022 do not resemble a circle, expected for the circular motion around the compact object, but likewise is more consistent with a line.
Similar behavior is observed for the astrometric flare from 28 July 2018.
The pattern detected in polarization are consistent with those shown in the right columns of Fig.~\ref{fig:QU_eta_flying_towards} and \ref{fig:QU_eta_flying_away}, i.e. hot spots at high inclinations and flying away from the orbital plane.
Likewise, the path of hot spots at inclinations closer to $90\degr$ is more squeezed, as compared to the nearly-circular one at low ($\sim0\degr$ or $180\degr$) inclinations (e.g. images of accreting BHs in fig.~9 of \citealt{Loktev2022}).
We suggest that the flares showing elongated paths (or elongated loops) in astrometry and/or polarimetry can correspond to hot spots passing the BH at high inclinations.

\begin{figure}
\centering
\includegraphics[width=0.8\linewidth]{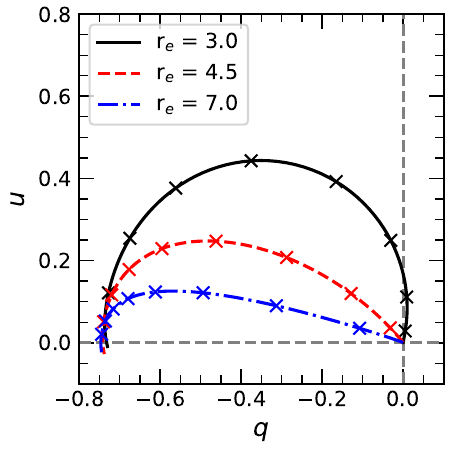}
\caption{Relative Stokes parameters (in fractional units) of the disk ring (in the equatorial plane $\eta=0$ and rotating in the counterclockwise direction $\sigma=90\degr$) with vertical field $B=B_z$, as a function of inclination for different ring radii. Zero polarization corresponds to the case of $i=0$ and increases, along the line, to $i=90\degr$. Inclinations $i=10\degr$, $20\degr$, $30\degr$...$80\degr$ are marked with crosses. For the inclinations $i>90\degr$ the curve is reflected around the $u=0$ axis.
}\label{fig:UQ_precessing_ring}
\end{figure}

If the extent of the hot spot becomes large, the PD drops, but does not generally become zero. 
For the case of the disk ring, the net polarization arises owing to the fact of the broken symmetry as a result of the combined GR and SR effects.
The net polarization of $\sim5$\% was observed in Sgr~A* \citep{EHT2024VII_Sgr}.
In Fig.~\ref{fig:UQ_precessing_ring} we show the net polarization (Stokes parameters) of the narrow disk ring at various radii, as a function of inclination.
All points intersect at the origin $(q,u)=(0,0)$ for $i=0$ and the polarization generally increases with increasing $i$.
The closer the disk ring is located to the BH, the higher is the net polarization and the more the PA deviates from the direction along the disk normal.
Hence, the $u$ parameter of the net polarization can be used to get information about the proximity of matter to the compact object.

The polarimetric signatures of the disk ring depend on the inclination, hence they are expected to change if the ring undergoes Lense-Thirring precession.
The precessing disk rings scenario has been considered in the context of the quasi-periodic eruptions \citep{Raj2021,Arcodia2021}.
In the course of the Lense-Thirring precession, the inclination to the disk changes in a periodic way within limits set by the observer's orientation relative to the BH spin and the misalignment angle between the BH spin and the plane of matter inflow.
Intrinsic dependence of the polarimetric signatures on the inclination in Fig.~\ref{fig:UQ_precessing_ring} can then be converted to the observed profiles simply by computing the inclination of the disk at each precession phase \citep{VPI2013}.

\begin{figure}
\centering
\includegraphics[width=0.8\linewidth]{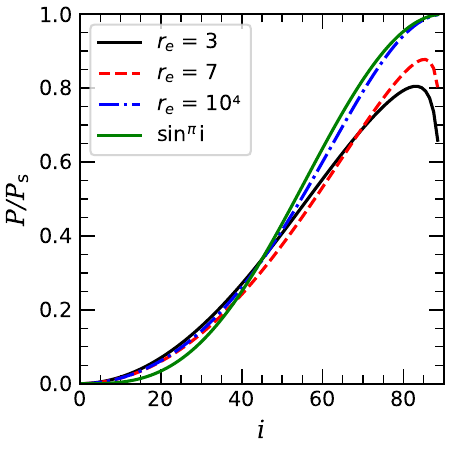}
\caption{PD as a function of inclination, scaled to its maximal value $P_{\rm s}$, for the case of toroidal magnetic field ($\phi'=90\degr$), equatorial counterclockwise rotation ($\eta=0$, $\sigma=90\degr$).
The case $r_{\rm e}=10^4$ shows the dependence in the absence of relativistic effects and can be well approximated by the $\sin^{\pi}{i}$. 
}\label{fig:PD_i}
\end{figure}

In Fig.~\ref{fig:PD_i} we show the dependence of the polarization degree on the inclination angle for the case of toroidal magnetic field $\phi'=90\degr$ and different radii.
The case of $r=10^4$ might be relevant to the outer parts of the jet, as the ratio of the poloidal and toroidal components of the field decreases linearly with distance in the standard \citet{BK1979} jet model.
At small viewing angles, the PD is small because positive contributions to polarization coming from azimuthal angles $\varphi$ and $\varphi+180\degr$ are (nearly) canceled by the negative contributions coming from the azimuthal angles $\varphi\pm90\degr$, for any $\varphi$.
Light bending and aberration alter the PD at high inclinations, where otherwise (for slow motions and no light bending) the contributions from different parts of the disk have nearly the same polarization angle, hence the PD reaches maximal value $P_{\rm s}$.
For small distances from the compact object we observe the effect of depolarization, caused by the different viewing angles of different parts of the disk; it is mostly pronounced at large inclinations.
A good approximation for the case of a high radius is given by the $\sin^{\pi}i$ function (green solid line in Fig.~\ref{fig:PD_i}).

For toroidal fields, the net polarization is aligned with the disk(/jet) axis for matter located at large distances.
Close to the BH, the net PA deviates from this direction by up to $12\degr$, achieved at $r_{\rm e}=3$.
The PA deviation weakly depends on the inclination.

\begin{figure}
   \centering
         \includegraphics[width=\linewidth]{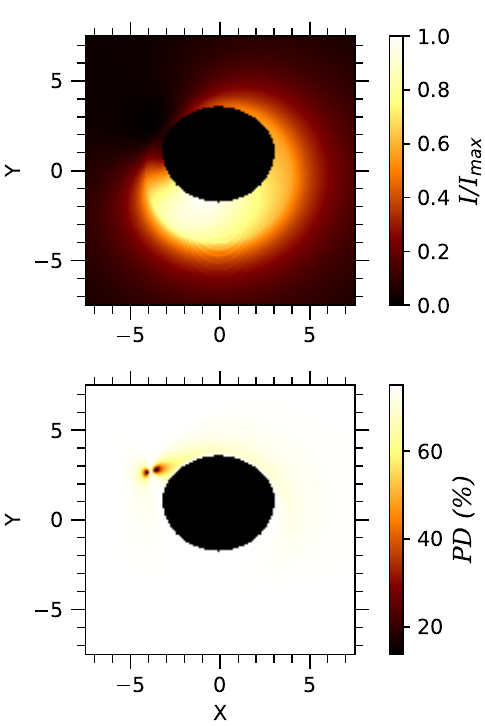}
   \caption{Color-coded image of the thick accretion disk around a black hole. In the upper panel the color refers to the monochromatic intensity, in the lower panel -- to PD. The drop of PD close to the critical point can be seen in the upper-left part of the image; otherwise, some depolarization is caused by the non-zero disk thickness. }
              \label{ThickBHDisc}
    \end{figure}

\subsection{Image of the geometrically thick disk}

Polarimetric images of two accreting supermassive BH environments recently obtained by the Event Horizon Telescope \citep{EHT2021VIII,EHT2024VII_Sgr} show a consistent picture of spiral polarization pattern that is dominated by the azimuthally symmetric structure.
The images correspond to the quiescent level of Sgr~A* and M~87* emission and both show the rotationally symmetric pattern of PA rotation.

In Fig.~\ref{ThickBHDisc} we plot the intensity (Eq.~\ref{eq:flux}) in the upper panel and the PD in the lower panel.
The disk is assumed to be geometrically thick, with $\eta=30\degr$ at the ISCO, that translates to the constant height of $h=1.73\rs$.
We split the disk into eleven horizontal layers (one layer in the disk mid-plane, five layers below and above it) and assume the elevation of matter above the mid-plane to be constant within each layer.
We verified that this resolution is sufficient for convergence purposes.
The observer inclination $i=30\degr$ and vertical magnetic field $B=B_z$ are assumed. 
We obtain the characteristic image of an asymmetric ring-like structure, similar to the observed ones in M~87* and Sgr~A*, as well as to the ones obtained in modeling of earlier works \citep{Narayan2021,Vincent2023,EHT2024VIII_Sgr}.
However, the area of the region with high intensity is larger in the case of the thick disk, as compared to the flat disk, and the shape of the BH shadow is squashed for the case of the thick disk.
Both effects are caused by the presence of matter above the equatorial plane, at azimuthal angles close to zero (i.e., between the observer and the BH).

The image of the disk is depolarized as compared to the results for the narrow, flat ring. 
The depolarization is caused by the substantial vertical thickness: emission coming from different layers of the disk has different rotation angles of polarization plane (see contours of PA rotation for different $\eta$ in Fig.~\ref{fig:chiGRSR_contours}), hence the net polarization is reduced.
The dark area  around coordinates $(-4,3)$ corresponds to the place where the disk is viewed face-on as a result of light bending and aberration effects.
For the vertical field, both intensity and PD decrease to zero here, and non-zero values appear as a result of the different position of this point for different disk layers.
For the direct comparison to the data, additional beam depolarization and internal Faraday rotation \citep[that is known to affect the polarization signatures in sub-mm range,][]{EHT2021VIII,EHT2024VII_Sgr,Wielgus2024} need to be considered.
The latter effect can be taken into account by assuming each layer between the emission point and the observer acts as a Faraday screen.

\section{Summary}
\label{sect:summary}

In this study, we have derived explicit analytical expressions for the transformation of polarization signatures for a fast-moving matter in Schwarzschild metric for the case of the synchrotron emission.
The rotation of polarization angle is caused by the joint effects of general and special relativity, as well as magnetic field orientation, which are considered separately.
We consider different cases of moving matter direction, its elevation above and below the equatorial plane and magnetic field orientation, and compute the resulting polarization angles as a function of azimuthal coordinate in the disk.
We further discuss potential applications of the developed formalism for the dynamic polarimetric signatures of spots orbiting around the compact object, as well as static images of geometrically thick accretion flows.
We find that a number of the observed signatures can be explained by the non-zero thickness of the disk, that was generally omitted in previous analytical results.
We highlight an important difference of the calculations presented in this work, as compared to the polarization signatures in Minkowski space.
The simple form of the derived expressions makes them attractive for the acceleration of the post-processing calculations, that can be done for direct comparison with observations.

\section*{Acknowledgments}
AV acknowledges support the Academy of Finland grant 355672.  
Nordita is supported in part by NordForsk.
Authors thank V. Loktev, J. Poutanen and M. Wielgus for the useful discussions and comments on the paper.

\bibliographystyle{aa}

\appendix

\section{Rotation of polarization angle}
\label{sect:appendixA}

Light bending and relativistic aberration alter the direction of the photon path and hence change the direction of its electric field orientation, as seen by the observer.
To determine the rotation of PA, we follow the method described in \cite{Loktev2022}.
For an arbitrary vector ${\bm l}$, not collinear with the photon propagation direction $\unit{k}$, we introduce the polarization basis, formed by two vectors orthogonal to each other and to the vector $\unit{k}$ as
\begin{equation}
    \mathcal{B}(\unit{k},\bm{l}) = \left\{ \hat {\bm e}_1^{k,l}, \hat {\bm e}_2^{k,l}\right\}=\left\{\frac{\bm{l}-(\unit{k}\cdot\bm{l})\unit{k}}{|\unit{k}\times\bm{l}|},\frac{\unit{k}\times\bm{l}}{|\unit{k}\times\bm{l}|}\right\}.
\end{equation}
The PA is given by the angle between the polarization plane of the photon and $\hat {\bm e}_1^{k,l}$, in the counterclockwise direction, as viewed by the observer intercepting the photon.
For any other vector $\bm{l}_2$, the new basis is $\mathcal{B}(\unit{k},\bm{l}_2)$, and the PA is given by the angle with respect to the new axis $\hat {\bm e}_1^{k,l_2}$.
The new basis is built around the same photon direction, hence the PAs are different by the angle of rotation between the bases, 
\begin{equation}
    \chi(\unit{k}, \bm{l}_2 \rightarrow \bm{l}) = \mathcal{B}(\unit{k}'_0, \unit{B}') \rightarrow \mathcal{B}(\unit{k}'_0, \unit{n}), 
\end{equation}
that can be computed as
\begin{equation}
    \tan \chi(\unit{k}, \bm{l}_2 \rightarrow \bm{l})= \frac{\unit{k}\cdot(\bm{l}\times\bm{l}_2)}{\bm{l}\cdot\bm{l}_2-(\unit{k}\cdot\bm{l})(\unit{k}\cdot\bm{l}_2)}.
    \label{eq:tanchi}
\end{equation}

For the synchrotron radiation of matter close to the BH, the photon is initially emitted along the direction $\unit{k}_0'$ with polarization orthogonal to the magnetic field lines $\unit{B}'$, the latter defines the basis $\mathcal{B}(\unit{k}'_0,\unit{B}')$.
Next, we shift to the description of the photon polarization plane orientation with respect to the normal to the plane of accreting matter $\unit{n}$.
Further, the relativistic aberration lead to the transition from $\unit{k}'_0$ to $\unit{k}_0$.
Finally, the light bending changes the photon vector $\unit{k}_0$ to $\unit{o}$.
Hence, we make the following steps between the bases:
\begin{equation}
     \mathcal{B}(\unit{k}'_0, \unit{B}') \rightarrow \mathcal{B}(\unit{k}'_0, \unit{n}) \rightarrow 
     \mathcal{B}(\unit{k}_0, \unit{n}) \rightarrow \mathcal{B}(\unit{o}, \unit{n}).
\end{equation}
Each step is associated with the rotation of PA by the angle between the bases and is associated with different effects: magnetic field orientation $\chi^{B}$, aberration $\chi^{\rm SR}$ and light bending $\chi^{\rm GR}$.
This can be explicitly written as
\begin{equation}
    \chi=\chi_0+\chi^{\rm{tot}}=\chi_0+\chi^{\rm{B}}+\chi^{\rm SR}+\chi^{\rm GR},
\end{equation}
where $\chi_0=90\degr$.

Rotation associated with the magnetic field orientation is computed as the shift 
$\mathcal{B}(\unit{k_0'}, \unit{B})\rightarrow \mathcal{B}(\unit{k_0'}, \unit{n})$.
Using Eq.~\ref{eq:tanchi}, $\chi^{\rm{B}}$ is computed as:
\begin{equation}\label{eq:PAmagnvector_appendix}
     \chi^{\rm{B}}=\arctan\left( \frac{\unit{k}^{\prime}_0\cdot(\unit{n}\times\unit{B'})}{\unit{n}\cdot\unit{B'}-(\unit{k}^{\prime}_0\cdot\unit{n})(\unit{k}^{\prime}_0\cdot\unit{B'})} \right).
\end{equation}
Explicit analytical expression for the considered geometry and matter motions is given in Eq.~\eqref{eq:chiB}.

The second and the third transformations have been described in \citet{Loktev2022} and we briefly mention them below for the sake of completeness.
The transformation $\mathcal{B}(\unit{k}'_0, \unit{n}) \rightarrow \mathcal{B}(\unit{k}_0, \unit{n})$ is done via three steps:
\begin{equation}
    \mathcal{B}(\unit{k}'_0, \unit{n}) \rightarrow \mathcal{B}(\unit{k}'_0, \unit{v}) \rightarrow \mathcal{B}(\unit{k}_0, \unit{v}) \rightarrow \mathcal{B}(\unit{k}_0, \unit{n}),
\end{equation}
where the second step introduces zero rotation, as the vectors $\unit{k}'_0$, $\unit{v}$ and $\unit{k}_0$ lie in the same plane.
For the considered case $\unit{v} \perp \unit{n}$, the rotation angle is obtained as
\begin{equation}
    \tan \chi^{\rm SR} = - \frac{\beta (\unit{k}_0 \cdot (\unit{v} \cross \unit{n})) (\unit{k}_0 \cdot \unit{n})}
                                {1 - (\unit{k}_0\cdot\unit{n})^2 - \beta (\unit{k}_0 \cdot \unit{v})}.
\end{equation}

The transformation $\mathcal{B}(\unit{k}_0, \unit{n}) \rightarrow \mathcal{B}(\unit{o}, \unit{n})$ is done via an additional step involving the radial vector direction $\unit{r}$:
\begin{equation}
    \mathcal{B}(\unit{k}_0, \unit{n}) \rightarrow \mathcal{B}(\unit{k}_0, \unit{r}) \rightarrow \mathcal{B}(\unit{o}, \unit{r}) \rightarrow \mathcal{B}(\unit{o}, \unit{n}),
\end{equation}
where, again, the second step introduces zero rotation, as the vectors $\unit{k}_0$, $\unit{r}$ and $\unit{o}$ belong to the same plane in Schwarzschild metric (photon trajectories lie in plane).
The resulting vector expression for the rotation angle is
    \begin{equation}
    \tan \chi^{\rm GR} = \cos\eta 
    \frac{(\unit{o} \cdot \unit{\varphi}) \Tilde{g}_1 - (\unit{k}_0 \cdot \unit{\varphi}) \Tilde{g}_2}{\Tilde{g}_1\Tilde{g}_2 + \cos^2\eta (\unit{o} \cdot \unit{\varphi})(\unit{k}_0 \cdot \unit{\varphi})},
\end{equation}
where
\begin{eqnarray}
    \Tilde{g}_1 &=& (\unit{r} \cdot \unit{n}) - (\unit{k}_0 \cdot \unit{n})(\unit{k}_0 \cdot \unit{r}), \\
    \Tilde{g}_2 &=& (\unit{r} \cdot \unit{n}) - (\unit{o} \cdot \unit{n})(\unit{o} \cdot \unit{r}).
\end{eqnarray}

\end{document}